\documentclass[prl,twocolumn,showpacs,amsmath,amssymb]{revtex4-1}

\usepackage{graphicx} 
\usepackage{dcolumn} 
\usepackage{multirow} 
\usepackage{hhline} 
\usepackage[table,usenames]{xcolor} 
\usepackage{bm} 
\usepackage[breaklinks=true]{hyperref} 

\begin{document}

\makeatletter
\def\flushboth{%
  \let\\\@normalcr
  \@rightskip\z@skip \rightskip\@rightskip
  \leftskip\z@skip
  \parindent 0em\relax}
\makeatother

\makeatletter
\def\CT@@do@color{\global\let\CT@do@color\relax\@tempdima\wd\z@\advance\@tempdima\@tempdimb\advance\@tempdima\@tempdimc\advance\@tempdimb\tabcolsep\advance\@tempdimc\tabcolsep\advance\@tempdima2\tabcolsep\kern-\@tempdimb\leaders\vrule\hskip\@tempdima\@plus 1fill\kern-\@tempdimc\hskip-\wd\z@ \@plus -1fill}
\makeatother

\definecolor{green}{HTML}{009900}
\definecolor{red}{HTML}{FF0000}
\definecolor{yellow}{HTML}{FFFF00}

\def\csg#1{{\color{red}\tt #1}}
\def\csw#1{{\color{green}\tt #1}}


\title{Accuracy and transferability of GAP models for tungsten}
\author{Wojciech J. Szlachta}
\affiliation{Engineering Laboratory, University of Cambridge, Trumpington Street, Cambridge, CB2 1PZ, UK}
\author{Albert P. Bart\'ok}
\affiliation{Engineering Laboratory, University of Cambridge, Trumpington Street, Cambridge, CB2 1PZ, UK}
\author{G\'abor Cs\'anyi}
\affiliation{Engineering Laboratory, University of Cambridge, Trumpington Street, Cambridge, CB2 1PZ, UK}

\date{\today} 

\begin{abstract}
We introduce  interatomic potentials for tungsten  in the bcc crystal phase  and its defects within the Gaussian Approximation Potential (GAP) framework, fitted to a database of first principles density functional theory (DFT)  calculations.  We investigate the performance of a sequence of models based on  databases of increasing coverage in configuration space and showcase our strategy of choosing representative small unit cells to train models that predict properties only observable using  thousands of atoms. The most comprehensive model  is then used to calculate  properties of the screw dislocation, including its structure, the Peierls barrier and the  energetics of the vacancy-dislocation interaction. All software and raw data are available at www.libatoms.org.
\end{abstract}


\maketitle

Tungsten is a hard, refractory metal with the highest melting point (3695 K) among metals, and its alloys are 
utilised in numerous  technological applications. 
The details of the atomistic processes behind the plastic behaviour of tungsten have been investigated for a long 
time and many interatomic potentials exist in the literature reflecting an evolution, over the 
past three decades, in their level of sophistication, starting with the Finnis-Sinclair (FS) potential \cite{doi:10.1080/01418618408244210}, embedded atom model (EAM) \cite{PhysRevB.29.6443}, various other FS/EAM 
parametrisations \cite{doi:10.1080/01418618708204464, doi:10.1080/09500839008206493, 0965-0393-22-1-015004, 0295-5075-26-8-005}, modified embedded atom models (MEAM) \cite{PhysRevB.46.2727, :/content/aip/journal/jap/78/1/10.1063/1.360661, PhysRevB.64.184102, 0953-8984-25-39-395502}  
and bond order potentials (BOP) \cite{PhysRevB.75.104119, :/content/aip/journal/jap/107/3/10.1063/1.3298466, Li201112}.
While some of these methods have been used to study other transition metals \cite{PhysRevB.38.3199, PhysRevB.54.6941, PhysRevB.69.094115}, there is renewed interest in modelling tungsten due to its many high 
temperature applications---e.g. it is one of the candidate materials for plasma facing components in the JET and 
ITER fusion projects \cite{1402-4896-2007-T128-027, 0741-3335-49-12B-S04, Pitts2013S48}.

\begin{figure}[t!]
\vspace{-0.33cm}
\hspace*{-0.5cm}
\large{}
\resizebox{10cm}{!}{\includegraphics{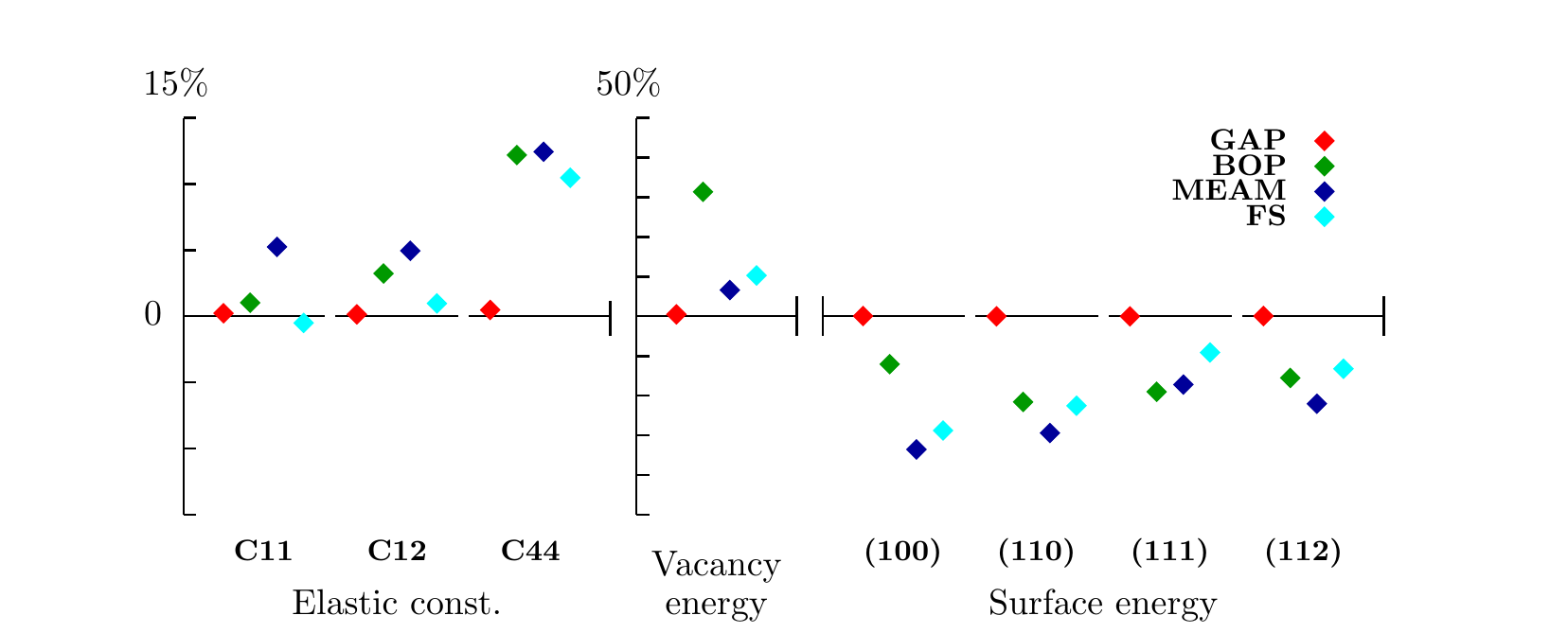}}
\normalsize{}
\vspace{-0.33cm}
\caption{Fractional error in elastic constants and defect energies calculated with various interatomic potentials, as compared to the target DFT values.}
\label{figure:elastic-constants-defects}
\end{figure}

A recurring problem with empirical potentials, due to the use of fixed functional forms with only 
a few adjustable parameters, is the lack of flexibility:  when fitted to reproduce  a given 
property, predictions for other properties can have large errors.
Figure~\ref{figure:elastic-constants-defects} shows the basic performance of BOP 
and MEAM, two of the more sophisticated potentials that reproduce the correct screw dislocation core 
structure, and also the simpler FS, all in comparison with density functional theory (DFT). While the figure emphasises fractional accuracy, we show the
corresponding absolute numerical values in
Table~\ref{table:elastic_constants_and_defects}. BOP is poor in 
describing the vacancy but is better at surfaces, whereas MEAM is the other way 
around. 
While this compromise can sometimes be made with good judgement for specific applications, many interesting 
properties, particularly those that determine the material behaviour at larger length scales, arise from the 
competition between different atomic scale processes, which therefore all need to be described equally well.
For example, dislocation pinning, depinning and climb involve both elastic properties,  
core structure, as well as the interaction of dislocations with  defects. One way to deal with this problem is to use 
multiple levels of accuracy as in QM/MM \cite{Bernstein2009} or to allow the parameters of the potential to vary 
in time and space \cite{Devita1997}.

\begin{table}
\begin{ruledtabular}
\begin{tabular}{ l c c c c c }
\\[-0.67em]
 & DFT & GAP & BOP & MEAM & FS \\
\\[-0.67em] \hline \\[-0.67em]
C11 [GPa] & 517 & 518 & 522 & 544 & 514 \\
C12 [GPa] & 198 & 198 & 205 & 208 & 200 \\
C44 [GPa] & 142 & 143 & 160 & 160 & 157 \\
\\[-0.67em] \hline \\[-0.67em]
vacancy energy [eV] & 3.27 & 3.29 & 4.30 & 3.49 & 3.61 \\[0.33em]
100 surface [eV/\r{A}$^2$] & 0.251 & 0.252 & 0.221 & 0.167 & 0.179 \\
110 surface [eV/\r{A}$^2$] & 0.204 & 0.204 & 0.160 & 0.144 & 0.158 \\
111 surface [eV/\r{A}$^2$] & 0.222 & 0.222 & 0.180 & 0.184 & 0.202 \\
112 surface [eV/\r{A}$^2$] & 0.216 & 0.216 & 0.182 & 0.168 & 0.187 \\
\\[-0.67em]
\end{tabular}
\end{ruledtabular}
\caption{Elastic constants and defect energies calculated with various interatomic potentials, and corresponding target DFT values.}
\label{table:elastic_constants_and_defects}
\end{table}

Here we describe a milestone in a research programme aimed at creating a potential that  circumvents the  problem of fixed functional forms. The purpose of the present work is twofold. Firstly, we showcase the power of the non-parametric database driven approach by constructing an accurate potential and using it to compute atomic scale properties  that are inaccessible to DFT due to computational expense. 
Secondly, while there has been vigorous activity recently in developing such models, most of the attention has 
been focussed on the interpolation method and the neighbourhood descriptors (e.g. neural networks \cite{PhysRevLett.98.146401, PhysRevLett.100.185501, PhysRevB.85.045439}, Shepherd interpolation \cite{:/content/aip/journal/jcp/100/11/10.1063/1.466801, Collins2002}, invariant polynomials \cite{doi:10.1021/jp048339l, :/content/aip/journal/jcp/122/4/10.1063/1.1834500, :/content/aip/journal/jcp/122/22/10.1063/1.1927529}, Gaussian processes \cite{PhysRevLett.104.136403, PhysRevB.87.184115, PhysRevB.88.054104, :/content/aip/journal/jcp/139/24/10.1063/1.4852182, PhysRevLett.108.058301}), rather 
less prominence was given to the question of how to construct suitable databases that ultimately determine the 
range of validity of the potential.
Our second goal is therefore to study what kinds of configurations need to be in a database so that given 
material properties are well reproduced. A larger database costs more to create and the resulting potential is 
slower, but can be expected to be more widely applicable, thus providing a tuneable tradeoff between 
transferability, accuracy and computational cost.

In our Gaussian Approximation Potential (GAP) framework \cite{PhysRevLett.104.136403, PhysRevB.87.184115}, the only uncontrolled approximation is the one essential to the idea of interatomic potentials: the 
total energy is written as a sum of atomic energies, 
\begin{align}
E = \sum_i\varepsilon(\mathbf{\hat q}_i),
\label{equation:ip}
\end{align}
with $\varepsilon$ 
a universal function of the atomic neighbourhood structure inside a finite
cutoff radius as represented by the descriptor vector $\mathbf{\hat q}_i$  for atom $i$ (defined below). This function is fitted to a database of DFT calculations using  Gaussian process regression \cite{mackay2003information, rasmussen2006gaussian} so, in general,  it is 
given by a linear combination of basis functions,
\begin{align}
\varepsilon(\mathbf{\hat q}) &= \sum_{j} \alpha_{j} K(\mathbf{\hat q}_{j}, \mathbf{\hat q})\equiv \mathbf{k}(\mathbf{\hat q})^T\bm{\alpha},
\label{equation:gp-mean} 
\end{align}
where the sum  over $j$ includes (some or all of) the configurations in the database, 
the vector of coefficients $\bm\alpha$ are given by linear algebra expressions (see
below and in \cite{PhysRevLett.104.136403}),
and the meaning of the covariance kernel $K$ is that of a similarity measure between different neighbour 
environments.

\begin{table}
\begin{ruledtabular}
\begin{tabular}{ l r }
\\[-0.67em]
DFT code & \begin{tabular}{@{}r@{}}CASTEP \cite{clark05-zkryst}\\(version 6.01)\end{tabular} \\
\\[-0.67em]
\hline
\\[-0.67em]
exchange-correlation functional & PBE \\
\\[-0.67em]
\hline
\\[-0.67em]
pseudopotential & \begin{tabular}{@{}r@{}}ultrasoft\\(valence 5s$^2$ 5p$^6$ 5d$^4$ 6s$^2$)\end{tabular} \\
\\[-0.67em]
\hline
\\[-0.67em]
plane-wave energy cutoff & $600$ $\text{eV}$ \\
\\[-0.67em]
\hline
\\[-0.67em]
maximum $k$-point spacing & $0.015$ $\text{\AA}^{-1}$ \\
\\[-0.67em]
\hline
\\[-0.67em]
electronic smearing scheme & Gaussian \\
\\[-0.67em]
\hline
\\[-0.67em]
smearing width & $0.1$ $\text{eV}$ \\
\\[-0.67em]
\hline \hline
\\[-0.67em]
atomic environment kernel & SOAP \\
\\[-0.67em]
\hline
\\[-0.67em]
\multicolumn{2}{c}{$
f_{cut} (r) = \begin{cases}
1 & 0 < r \leq (r_{\text{cut}} - r_{\Delta}) \\
\frac{1}{2} \cos( 1 + \pi \frac{r - r_{\text{cut}} + r_{\Delta}}{r_{\Delta}} ) & (r_{\text{cut}} - r_{\Delta}) < r \leq r_{\text{cut}} \\
0 & r_{\text{cut}} < r
\end{cases}
$} \\
\\[-0.67em]
\hline
\\[-0.67em]
\multicolumn{2}{c}{$
\begin{aligned}
\phi_{n} (r) &= \exp[-(r - r_{\text{cut}}n/n_\textrm{max})^{2}/2\sigma^2_\textrm{atom}] \\
\\[-0.67em]
S_{nn^\prime}&=\int_0^{r_\textrm{cut}}\!\!\!\!\!\!\!\!\!\!dr \,r^2 \phi_n(r)\phi_{n^\prime}(r)
\qquad \mathbf{S}=\mathbf{U}^T\mathbf{U}
\\[-0.67em]
\\[-0.67em]
\\[-0.97em]
g_n (r) &= \sum_{{n^\prime}} (\mathbf{U}^{-1})_{n {n^\prime}} \phi_{n^\prime} (r) 
\end{aligned}
$} \\
\\[-0.67em]
\hline
\\[-0.67em]
$r_{\text{cut}}$ & $5.0$ $\text{\AA}$ \\
$r_{\Delta}$ & $1.0$ $\text{\AA}$ \\
\\[-0.67em]
\hline
\\[-0.67em]
$\sigma_{\nu}^{\text{(energy)}}$ & $0.0001$ $\text{eV/atom}$ \\
$\sigma_{\nu}^{\text{(force)}}$ & $0.01$ $\text{eV/\AA}$ \\
$\sigma_{\nu}^{\text{(virial)}}$ & $0.01$ $\text{eV/atom}$ \\
\\[-0.67em]
\hline
\\[-0.67em]
$\sigma_{w}$ & $1.0$ $\text{eV}$ \\
\\[-0.67em]
\hline
\\[-0.67em]
$\sigma_{\text{atom}} $ & $0.5$ $\text{\AA}$ \\
\\[-0.67em]
\hline
\\[-0.67em]
$\xi$ & $4$ \\
\\[-0.67em]
\hline
\\[-0.67em]
$n_{\text{max}}$ & $14$ \\
$l_{\text{max}}$ & $14$ \\
\\[-0.67em]
\hline
\\[-0.67em]
GAP software version & \texttt{df1c4d9} \\
\\[-0.67em]
\end{tabular}
\end{ruledtabular}
\caption{DFT parameters used to generate training data and GAP model parameters.}
\label{table:GAP}
\end{table}

The expression for the coefficients $\alpha_j $---normally
simple in
Gaussian process regression---is more complicated in our case because the
quantum mechanical input data we can calculate is not a set of values of the atomic energy function that we are trying to fit. Rather, the total energy of a configuration is a sum of many atomic energy function values, and the forces and
stresses, which are also available analytically through the Hellmann-Feynman theorem, are sums of partial derivatives of the atomic energy function. The detailed derivation of the formulas shown below is in \cite{NIPS2005_543, 2010PhDT, 2013PhDT}. Let us collect all
the input data values (total energies, force and stress components) into
the vector $\mathbf{y}$ with $D$ components in total and denote by $\mathbf{y}^\prime$
the $N$ {\em
unknown} atomic energy values corresponding to all the atoms that appear
in all the input configurations. We construct a linear operator $\mathbf{L}$ that describes the relationship between them through $\mathbf{y} = \mathbf{L}^T \mathbf{y}^\prime$.
For data values that represent total energies, the corresponding rows of
$\mathbf{L}$ have just 0s and
1s as their elements, but for forces and stresses, the entries are differential
operators such as  $\partial/\partial x_i$ corresponding to the force
on atom $i$ with cartesian $x$ coordinate $x_i$. Writing $K_{ij} \equiv K(\mathbf{\hat q}_i,\mathbf{\hat q}_j)  $ for the element of the covariance matrix $\mathbf{K}_{NN}$ corresponding to atoms $i$ and $j$, the covariance matrix of size $D\times D$ of the observed data is,        
\begin{align}
\mathbf{K}_{DD} = \mathbf{L}^T \mathbf{K}_{NN} \mathbf{L},
\end{align}
where the differential operators in $\mathbf{L}$ act on the covariance function
$K$ that defines $\mathbf{K}_{NN}$. In our applications, $N$ can exceed  a hundred thousand, and therefore working with $N\times N$ matrices would be computationally very expensive. Because many atomic environments in our dataset are highly similar to one another, it is plausible that many fewer than $N$ atoms could be chosen to efficiently represent the range of neighbour environments. We choose $M$ representative atoms from the full set of $N$ atoms that appear in all the input configurations (typically with $M \ll N$), and denote the square covariance matrix between the $M$ representative atoms by $\mathbf{K}_{MM}$ and the rectangular
covariance matrix between the $M$ representative atoms and all the $N$ atoms by $\mathbf{K}_{MN}$ (with $\mathbf{K}_{NM} = \mathbf{K}_{MN}^T$). The expression for the vector of coefficients in equation \ref{equation:gp-mean} is then,
\begin{align}
\bm{\alpha} = [\mathbf{K}_{MM}+\mathbf{K}_{MN}\mathbf{L}\mathbf{\Lambda}^{-1}
\mathbf{L}^T\mathbf{K}_{NM}]^{-1}\mathbf{K}_{MN}\mathbf{L}\mathbf{\Lambda}^{-1}\mathbf{y},
\label{eq:alpha}
\end{align}
with 
\begin{align}
\mathbf{\Lambda} = \sigma^2_\nu \mathbf{I},
\end{align}
where the parameter $\sigma_\nu$ represents the tolerance (or expected error)\
in fitting the input data. It could be a single constant,  but in practice we found it essential to use different
tolerance values corresponding to the different kinds of input data, so that the
$\mathbf{\Lambda}$ matrix is still diagonal, but has different values corresponding
to total energies, forces and stresses as they appear in the data vector
$\mathbf{y}$.  Although one
might initially expect zero error in \textit{ab initio} input data, this
is not actually the case due to convergence parameters in the electronic
structure calculation. A further source of error in the fit is the  uncontrolled
approximation of equation \eqref{equation:ip}, i.e. writing the total
energy as a sum of local atomic energies.
The
numerical values we use are shown in Table~\ref{table:GAP}.
They are based on convergence tests of  the DFT calculation carried out on example configurations. 

We note the following remarks about the  expression in~\eqref{eq:alpha}. The quantum mechanically not defined and therefore unknown atomic energies for the input configurations, $\mathbf{y}^\prime$, do not appear. The number of components in the coefficient
vector ${\bm\alpha}$ is $M$, so the sum in equation \eqref{equation:gp-mean}
is over the $M$ representative configurations.  The cost of calculating  ${\bm \alpha}$ is dominated by operations which scale like $O(NM^2)$, so it can be significantly reduced by choosing $M$ to be smaller and accepting a
reduced  accuracy of the fit. After  the fit is made the coefficient
vector  ${\bm\alpha}$ stays fixed, and the evaluation of the potential 
is accomplished by the vector dot product in \eqref{equation:gp-mean} with
most of the work going towards computing the vector $\mathbf{k }$ for each new configuration, and thus scaling like $O(M)$. The $M$
representative atoms can be chosen randomly, but we found it beneficial to
employ the  k-means clustering algorithm to choose the representative configurations.   

We now turn to the specification of the kernel function. We use the ``smooth overlap of atomic positions'' (SOAP) 
kernel \cite{PhysRevB.87.184115}, 
\begin{align}
K_{ij} = \sigma^2_w|\mathbf{\hat q}_i\cdot\mathbf{\hat q}_j|^{\xi}
\label{eq:covariance}
\end{align}
where the exponent $\xi$ is a positive integer parameter whose role is to ``sharpen''
the selectivity of the similarity measure, and $\sigma_w$ is an overall
scale factor.
Note that for the special choice of $\xi=1$, the Gaussian process regression fit is equivalent to simple linear regression, and so potential energy expression in~\eqref{equation:gp-mean} simplifies  to $\varepsilon(\mathbf{\hat q}) = \left(\sigma_w^2\sum_j {\alpha}_j \mathbf{\hat q}_j\right)\cdot \mathbf{\hat q} $, in which the term in parentheses can be precomputed once and for all. Unfortunately we found that such a linear fit  significantly limits the attainable accuracy of the potential.

\begin{table*}
\begin{ruledtabular}
\begin{tabular}{ c l l | c || c | c | c | c | c || c | c}
\footnotetext[0]{\flushboth\footnotemark[1]~Time on a single CPU core of
Intel Xeon E5-2670 2.6GHz, \footnotemark[2]~RMS error, \footnotemark[3]~formation
energy error, \footnotemark[4]~RMS error of Nye tensor over the 12 atoms
nearest the dislocation core, cf. Figure~\ref{figure:peierls-barrier}.\\[-1em]}
\\[-0.67em] 
\parbox[c]{0.07\textwidth}{$ $} &
\parbox[c][7em][b]{0.07\textwidth}{Database:} &
\parbox[c]{0.34\textwidth}{$ $} &
\parbox[c]{0.05\textwidth}{\rotatebox[origin=c]{90}{$\,$ \begin{tabular}{@{}c@{}}Computational\\cost\footnotemark[1]
[ms/atom]\end{tabular}}} &
\parbox[c]{0.05\textwidth}{\rotatebox[origin=c]{90}{$\,$ \begin{tabular}{@{}c@{}}Elastic\\constants\footnotemark[2]{}
[GPa]\end{tabular}}} &
\parbox[c]{0.05\textwidth}{\rotatebox[origin=c]{90}{$\,$ \begin{tabular}{@{}c@{}}Phonon\\spectrum\footnotemark[2]
[THz]\\[-0.67em]{}\end{tabular}}} &
\parbox[c]{0.05\textwidth}{\rotatebox[origin=c]{90}{$\,$ \begin{tabular}{@{}c@{}}Vacancy\\formation\footnotemark[3]
[eV]\\[-0.67em]{}\end{tabular}}} &
\parbox[c]{0.05\textwidth}{\rotatebox[origin=c]{90}{$\,$ \begin{tabular}{@{}c@{}}Surface
energy\footnotemark[2]\\{} [eV/\r{A}$^2$]\\[-0.67em]{}\end{tabular}}} &
\parbox[c]{0.05\textwidth}{\rotatebox[origin=c]{90}{$\,$ \begin{tabular}{@{}c@{}}Dislocation\\structure\footnotemark[4]
[$\text{\r{A}}^{-1}$]\\[-0.67em]{}\end{tabular}}} &
\parbox[c]{0.05\textwidth}{\rotatebox[origin=c]{90}{\begin{tabular}{@{}c@{}}Dislocation-vacancy\\binding
energy [eV]\\[-0.67em]{}\end{tabular}}} &
\parbox[c]{0.05\textwidth}{\rotatebox[origin=c]{90}{\begin{tabular}{@{}c@{}}Peierls
barrier\\{}[eV/b]\\[-0.67em]{}\end{tabular}}} \\
\hhline{---|-||-|-|-|-|-||~|~} 
GAP$_{1}$ : &
\multicolumn{2}{l |}{\begin{tabular}{@{}l@{}}\\[-0.67em]2000 $\times$ primitive
unit cell\\with varying lattice vectors\\[-0.67em]{}\end{tabular}} &
24.70 &
\cellcolor{green!25}0.623 &
\cellcolor{red!25}\emph{0.583} &
\cellcolor{red!25}\emph{2.855} &
\cellcolor{red!25}\emph{0.1452} &
\cellcolor{red!25}\emph{0.0008} &
&
\\
\hhline{---|-||-|-|-|-|-||~|~} 
GAP$_{2}$ : &
GAP$_{1}$ + &
\begin{tabular}{@{}l@{}}\\[-0.67em]60 $\times$ 128-atom unit cell\\[-0.67em]{}\end{tabular}
&
51.05 &
\cellcolor{green!25}0.608 &
\cellcolor{green!25}0.146 &
\cellcolor{red!25}\emph{1.414} &
\cellcolor{red!25}\emph{0.1522} &
\cellcolor{red!25}\emph{0.0006} &
&
\\
\hhline{---|-||-|-|-|-|-||~|~} 
GAP$_{3}$ : &
GAP$_{2}$ + &
\begin{tabular}{@{}l@{}}\\[-0.67em]vacancy in: 400 $\times$ 53-atom unit
cell, \\20 $\times$ 127-atom unit cell\\[-0.67em]{}\end{tabular} &
63.65 &
\cellcolor{green!25}0.716 &
\cellcolor{green!25}0.142 &
\cellcolor{green!25}0.018 &
\cellcolor{red!25}\emph{0.0941} &
\cellcolor{yellow!25}\emph{0.0004} &
&
\\
\hhline{---|-||-|-|-|-|-||-|-} 
GAP$_{4}$ : &
GAP$_{3}$ + &
\begin{tabular}{@{}l@{}}\\[-0.67em]$(100)$, $(110)$, $(111)$, $(112)$ surfaces\\180
$\times$ 12-atom unit cell\\{}\\[-0.67em]$(110)$, $(112)$ gamma surfaces\\6183
$\times$ 12-atom unit cell\\[-0.67em]{}\end{tabular} &
86.99 &
\cellcolor{green!25}0.581 &
\cellcolor{green!25}0.138 &
\cellcolor{green!25}0.005 &
\cellcolor{green!25}0.0001 &
\cellcolor{green!25}0.0002 &
\cellcolor{yellow!25}\emph{-0.960} &
\cellcolor{green!25}\emph{0.108} \\
\hhline{---|-||-|-|-|-|-||-|-} 
GAP$_{5}$ : &
GAP$_{4}$ + &
\begin{tabular}{@{}l@{}}\\[-0.67em]vacancy in: $(110)$, $(112)$ gamma surface\\750
$\times$ 47-atom unit cell\\[-0.67em]{}\end{tabular} &
93.86 &
\cellcolor{green!25}0.865 &
\cellcolor{green!25}0.126 &
\cellcolor{green!25}0.011 &
\cellcolor{green!25}0.0001 &
\cellcolor{green!25}0.0002 &
\cellcolor{green!25}-0.774 &
\cellcolor{yellow!25}\emph{0.154} \\
\hhline{---|-||-|-|-|-|-||-|-} 
GAP$_{6}$ : &
GAP$_{5}$ + &
\begin{tabular}{@{}l@{}}\\[-0.67em]$\frac{1}{2} \langle 111 \rangle$ dislocation
quadrupole\\100 $\times$ 135-atom unit cell\\[-0.67em]{}\end{tabular} &
93.33 &
\cellcolor{green!25}0.748 &
\cellcolor{green!25}0.129 &
\cellcolor{green!25}0.015 &
\cellcolor{green!25}0.0001 &
\cellcolor{green!25}0.0001 &
\cellcolor{green!25}-0.794 &
\cellcolor{green!25}0.112 \\
\end{tabular}
\end{ruledtabular}
\caption{Summary of the databases for six GAP models, in order of increasing breadth in the types of configurations they contain, together with the performance of the corresponding potentials with respect to key properties. The colour of the cells indicates a subjective judgement of performance: unacceptable (red), usable (yellow), good (green). The first five properties  can be checked against DFT directly and so we report errors, but calculation of the last two properties are in large systems, so we report the values, converged with system size. The configurations are collected using Boltzmann sampling, for more details on the databases leading to the models see the supplementary
information.}
\label{table:training-database}
\end{table*}

\begin{table}
\begin{ruledtabular}
\begin{tabular}{ l c || c | c | c | c | c | c || c }
\\[-0.67em] 
$ $ & $ $ & \multicolumn{6}{c}{\begin{tabular}{@{}l@{}}\\[-0.67em]Database:\\[-0.67em]{}\end{tabular}} \\
\hhline{~~||~~~~~~||~}
& & 1 & 2 & 3 & 4 & 5 & 6 & \begin{tabular}{@{}c@{}}\\[-0.67em]Total\\$M$\\[-0.67em]{}\end{tabular} \\
\hhline{--||-|-|-|-|-|-||-}
\begin{tabular}{@{}l@{}}\\[-0.67em]GAP$_{1}$\\[-0.67em]{}\end{tabular} & & 2000 & \multicolumn{5}{c||}{$ $} & 2000 \\
\hhline{--||-|-|~~~~||-}
\begin{tabular}{@{}l@{}}\\[-0.67em]GAP$_{2}$\\[-0.67em]{}\end{tabular} & & 814 & 3186 & \multicolumn{4}{c||}{$ $} & 4000 \\
\hhline{--||-|-|-|~~~||-}
\begin{tabular}{@{}l@{}}\\[-0.67em]GAP$_{3}$\\[-0.67em]{}\end{tabular} & & 366 & 1378 & 4256 & \multicolumn{3}{c||}{$ $} & 6000 \\
\hhline{--||-|-|-|-|~~||-}
\begin{tabular}{@{}l@{}}\\[-0.67em]GAP$_{4}$\\[-0.67em]{}\end{tabular} & & 187 & 617 & 1890 & 6306 & \multicolumn{2}{c||}{$ $} & 9000 \\
\hhline{--||-|-|-|-|-|~||-}
\begin{tabular}{@{}l@{}}\\[-0.67em]GAP$_{5}$\\[-0.67em]{}\end{tabular} & & 158 & 492 & 1604 & 5331 & 2415 & $ $ & 10000 \\
\hhline{--||-|-|-|-|-|-||-}
\begin{tabular}{@{}l@{}}\\[-0.67em]GAP$_{6}$\\[-0.67em]{}\end{tabular} & & \parbox[c]{0.05\textwidth}{140} & \parbox[c]{0.05\textwidth}{450} & \parbox[c]{0.05\textwidth}{1500} & \parbox[c]{0.05\textwidth}{4874} & \parbox[c]{0.05\textwidth}{2211} & \parbox[c]{0.05\textwidth}{825} &  \parbox[c]{0.05\textwidth}{10000} \\
\end{tabular}
\end{ruledtabular}
\caption{Number of representative atomic environments in each database of the six GAP models. The rows represent the successive GAP models and the columns
represent the configuration types in the databases, grouped according to which GAP model first incorporated them. The allocations shown are based on k-means
clustering. The rightmost column shows the total number of representative
atoms in each GAP model ($M$).}
\label{table:sparse_points}
\end{table}

  The elements of the descriptor vector $\mathbf{\hat q}$ are constructed as follows. The  environment of the \(i\)th atom is characterised by the  \emph{atomic neighbourhood density}, which we define as
\begin{align}
\rho_i(\mathbf{r}) &= \sum_j e^{- |\mathbf{r}-\mathbf{r}_{ij}|^2/2\sigma^2_\textrm{atom}} f_\mathrm{cut}(|\mathbf{r}_{ij}|)\\
&= \sum_{n < n_\textrm{max} \atop {l < l_\textrm{max} \atop |m| \leq l }} c^i_{nlm} g_n(|\mathbf{r}|)Y_{lm}(\mathbf{\hat r})\nonumber
\end{align}
where $\mathbf{r}_{ij}$ are the vectors pointing to the neighbouring atoms,
$\sigma_\textrm{atom}$ is a parameter corresponding to the ``size'' of atoms, $f_\textrm{cut}$ is a smooth cutoff function with compact support, and the  expansion on the second line uses spherical harmonics and a set of orthonormal radial basis functions,
$g_n$, with $n$, $l$ and $m$ the usual integer indices. The elements of the descriptor vector $\mathbf{\hat q}$ are then,
\begin{align}
\mathbf{q}_i = \left\{\sum_m (c^i_{nlm})^*c^i_{n' lm} \right\}_{nn'l},\quad \mathbf{\hat q}_i &= \mathbf{q}_i / |\mathbf{q}_i|  
\end{align} 
Values for the all the parameters and other necessary formulas are given in
Table~\ref{table:GAP}.
The orthonormal radial basis is obtained from a set of equispaced Gaussians by Cholesky factorisation of their overlap matrix.

The SOAP kernel is special because it is not only invariant with respect to relabelling of atoms and  rotation of 
either neighbour environment, but it is also  faithful  in the sense that $K$ only takes the value of unity when the 
two neighbourhoods are identical. This is because  it is directly proportional  to the overlap of the\ atomic 
neighbourhood densities, integrated over all three dimensional rotations $\hat R$,
\begin{align}
K_{ij} \propto \left| \int\!\! d\hat R \left| \int \!\! d\mathbf{r} \rho_i(\mathbf{r})\rho_j(\hat R\mathbf{r})\right|^2 \right|^\xi.
\end{align}
The SOAP kernel is therefore also manifestly smooth and slowly varying in Cartesian space, just as we know the true Born-Oppenheimer potential energy surface to be, away from electronic energy level crossings and quantum phase transitions. The entire GAP framework, including the choice of descriptor and the kernel, is designed so that its parameters are easy to set and the final potential
is not very sensitive to the exact values. Some are physically motivated
and stem from either the properties of the quantum mechanical potential energy surface
($r_\textrm{cut}$, $\sigma_w$, $\sigma_\textrm{atom}$) or the input data
(e.g. $\sigma_\nu$), while others are
convergence parameters and are set by a tradeoff between accuracy and computational
cost ($n_\textrm{max}$, $l_\textrm{max}$, $M$). We include in the supplementary information a table demonstrating convergence
of the fitted potential as a function of $n_\textrm{max}$, $l_\textrm{max}$,
and $r_\textrm{cut}$. By far the most ``arbitrary''
part of the potential is thus the set of configurations chosen to  comprise
the training database. \

Since the potential interpolates the atomic energy in the space of neighbour environments,  we need good 
coverage of \emph{relevant} environments in the database. We  therefore need to start by deciding what material 
properties we wish to study and what are the corresponding neighbour environments. Our strategy is to define, 
for each material property, a set of representative \emph{small unit cell} configurations that are amenable to accurate first 
principles calculation. In Table~\ref{table:training-database} we show the performance with respect to key material properties of six models, each fitted 
to a database that contains the configurations indicated on the left, in addition to all the configurations of the 
preceding one. In particular, as proposed by Vitek \cite{doi:10.1080/14786436908217784, doi:10.1080/14786437008238490, doi:10.1080/14786430310001611644}, the structure of $\frac{1}{2} \langle 111 \rangle$ screw dislocations in bcc transition metals can be rationalised in terms of the strictly planar gamma 
surface concept, and therefore we use gamma surfaces in the database to ensure the coverage of  neighbour 
environments found near the dislocation core. Where the dislocation structure is very far from correct, the numerical 
performance metric on it has been omitted. The table shows that, broadly speaking, the small representative unit 
cells are necessary and also sufficient to obtain each property accurately, so the GAP model interpolates well but 
does not extrapolate to completely new kinds of configurations. Adding new configurations never compromises 
the accuracy of previously incorporated properties.
For information, Table~\ref{table:sparse_points} shows the results of the
automatic allocation of
the representative atoms in each GAP model to the various types of configurations.

\begin{figure}[t!]
\vspace{-0.5cm}
\hspace*{-0.5cm}
\large{}
\resizebox{10.0cm}{!}{\includegraphics{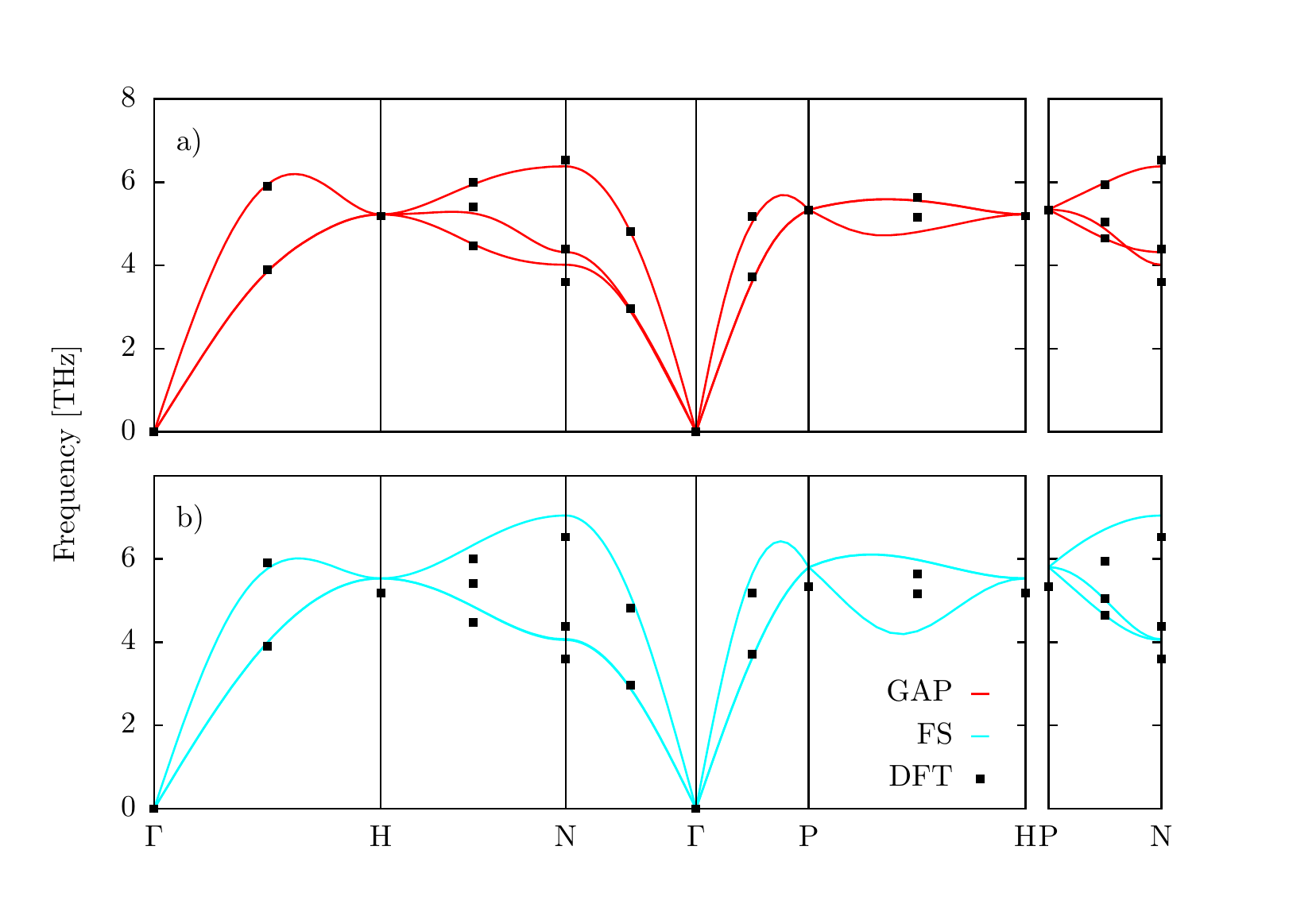}}
\normalsize{}
\vspace{-0.5cm}
\caption{Phonon spectrum of bcc tungsten calculated using GAP and FS potentials, and some  reference DFT values.}
\label{figure:phonons}
\end{figure}

 We also show the performance of the final GAP$_6$ model on 
Figure~\ref{figure:elastic-constants-defects} and omit the subscript from now. The phonon spectrum of the GAP model is shown in Figure~\ref{figure:phonons} along with that of the DFT  and FS.  There is clear improvement
with respect to the analytical model, but remaining deficiencies are also
apparent. Strategies to enhance the training database in order to improve the description of phonons is an important future direction of study.

\begin{figure}[t!]
\vspace{-0.5cm}
\hspace*{-0.5cm}
\large{}
\resizebox{10cm}{!}{\includegraphics{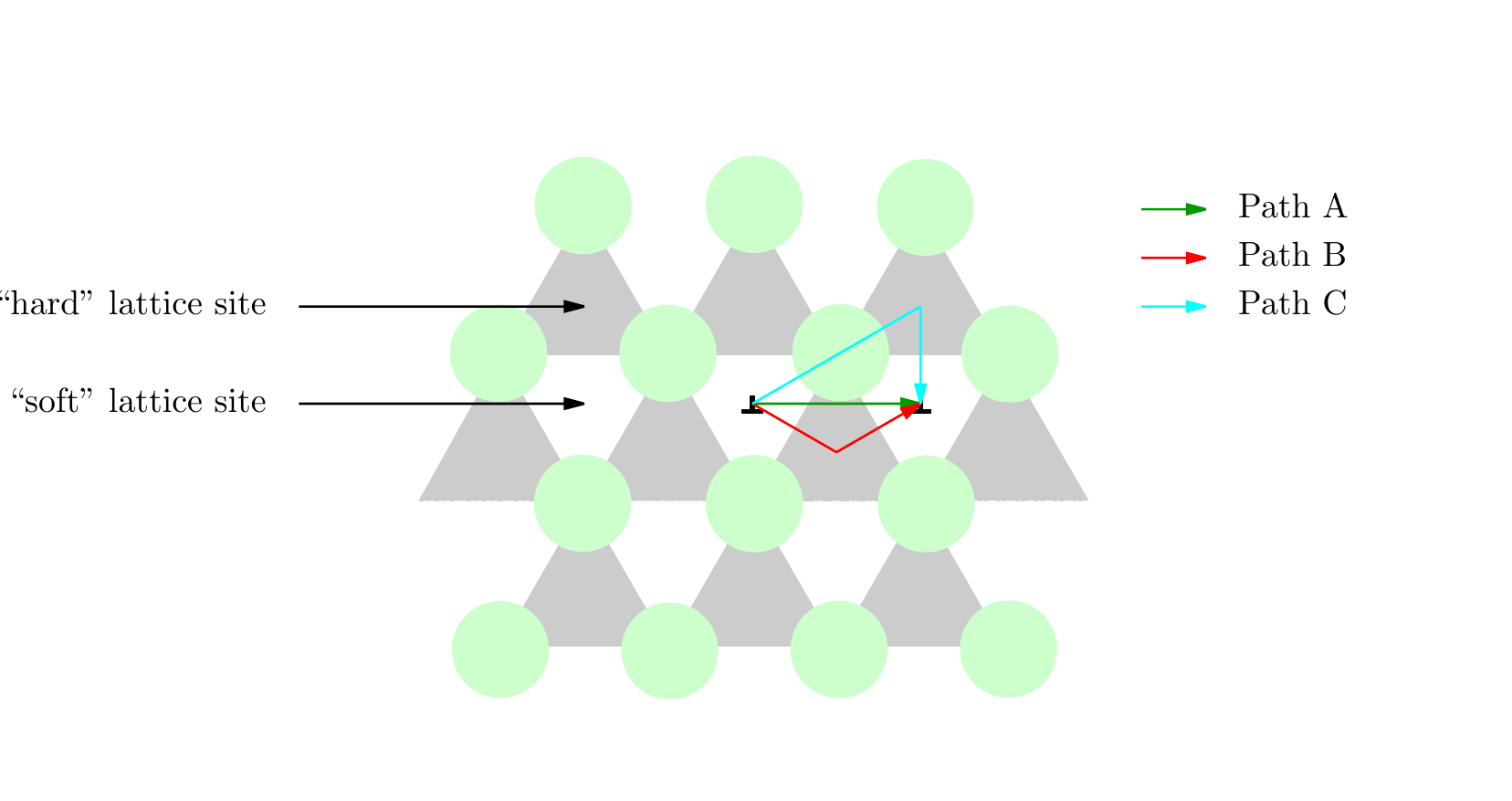}}
\normalsize{}
\vspace{-0.67cm}
\caption{Representation of the three different initial transition paths for
the Peierls barrier calculation. Path A corresponds to the linear interpolation directly from the initial to the final state, whereas paths B and C are the two distinct linear interpolations that include a potential meta-stable state (corresponding to the ``hard'' structure of the dislocation core) at reaction coordinate $r=0.5$.}
\label{figure:peierls-barrier-cell}
\end{figure}

\begin{figure}[t!]
\vspace{-0.5cm}
\hspace*{-0.5cm}
\large{}
\resizebox{10.0cm}{!}{\includegraphics{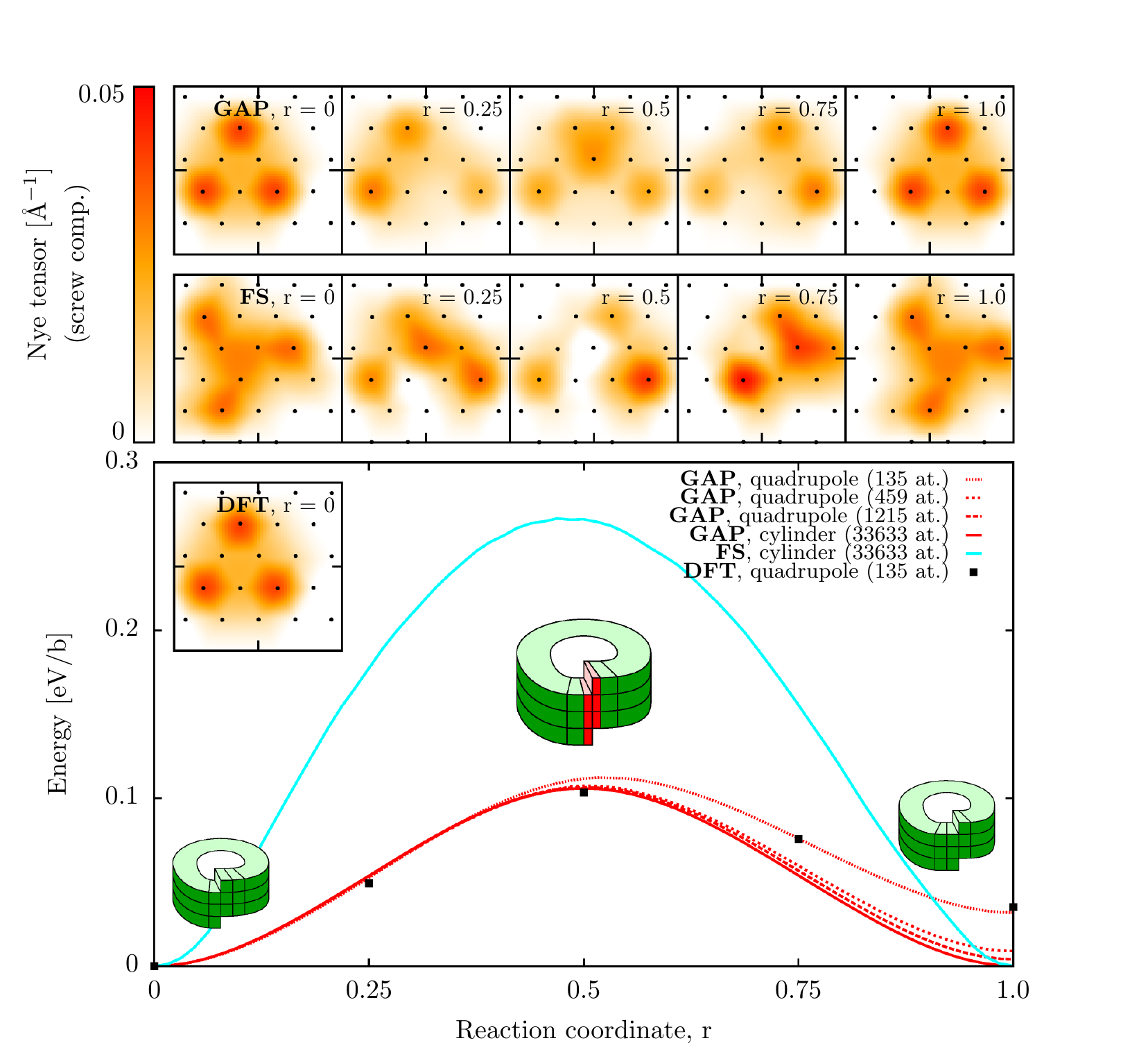}}
\normalsize{}
\vspace{-0.33cm}
\caption{Top:\ the structure of the screw dislocation along the minimum energy path as it glides; bottom: Peierls barrier evaluated using GAP and FS potentials, along with single point checks with DFT in the 135 atom quadrupole arrangement.}
\label{figure:peierls-barrier}
\end{figure}

We now investigate the properties of the $\frac{1}{2} \langle 111 \rangle$ screw dislocation further by calculating the 
Peierls barrier using a transition state searching implementation of the string method \cite{PhysRevB.66.052301, :/content/aip/journal/jcp/126/16/10.1063/1.2720838}. Three different initial transition paths, shown in Figure~\ref{figure:peierls-barrier-cell}, are used to explore the existence of the metastable state corresponding to a ``hard'' core structure \cite{PhysRevB.54.6941, PhysRevLett.84.1499, PhysRevB.68.014104, 0953-8984-25-8-085702}. We find that 
the ``hard'' core is not even locally stable in tungsten---starting geometry optimisation from there results in the 
dislocation line migrating to a neighbouring lattice site, corresponding to the ``soft'' core configuration. All three initial transition 
paths converge to the same minimum energy pathway (MEP), shown in Figure~\ref{figure:peierls-barrier},  with 
no ``hard'' core transition state. For large enough systems, the MEP is independent of the boundary conditions: 
the ``quadrupole'' calculations contained two oppositely directed dislocations in periodic boundary conditions, 
while the ``cylinder'' configurations had a single dislocation with fixed far field boundary conditions.
For comparison we also plot the MEP of the Finnis-Sinclair model, and show the corresponding core structures 
using Nye tensor maps \cite{Hartley20051313, doi:10.1080/14786430600660849}. For the smallest periodic 135 
atom model, we computed the energies at five points along the MEP using DFT to verify that the GAP model is 
indeed accurate for these configurations.

\begin{figure}[b!]
\hspace*{-0.5cm}
\large{}
\resizebox{10.0cm}{!}{\includegraphics{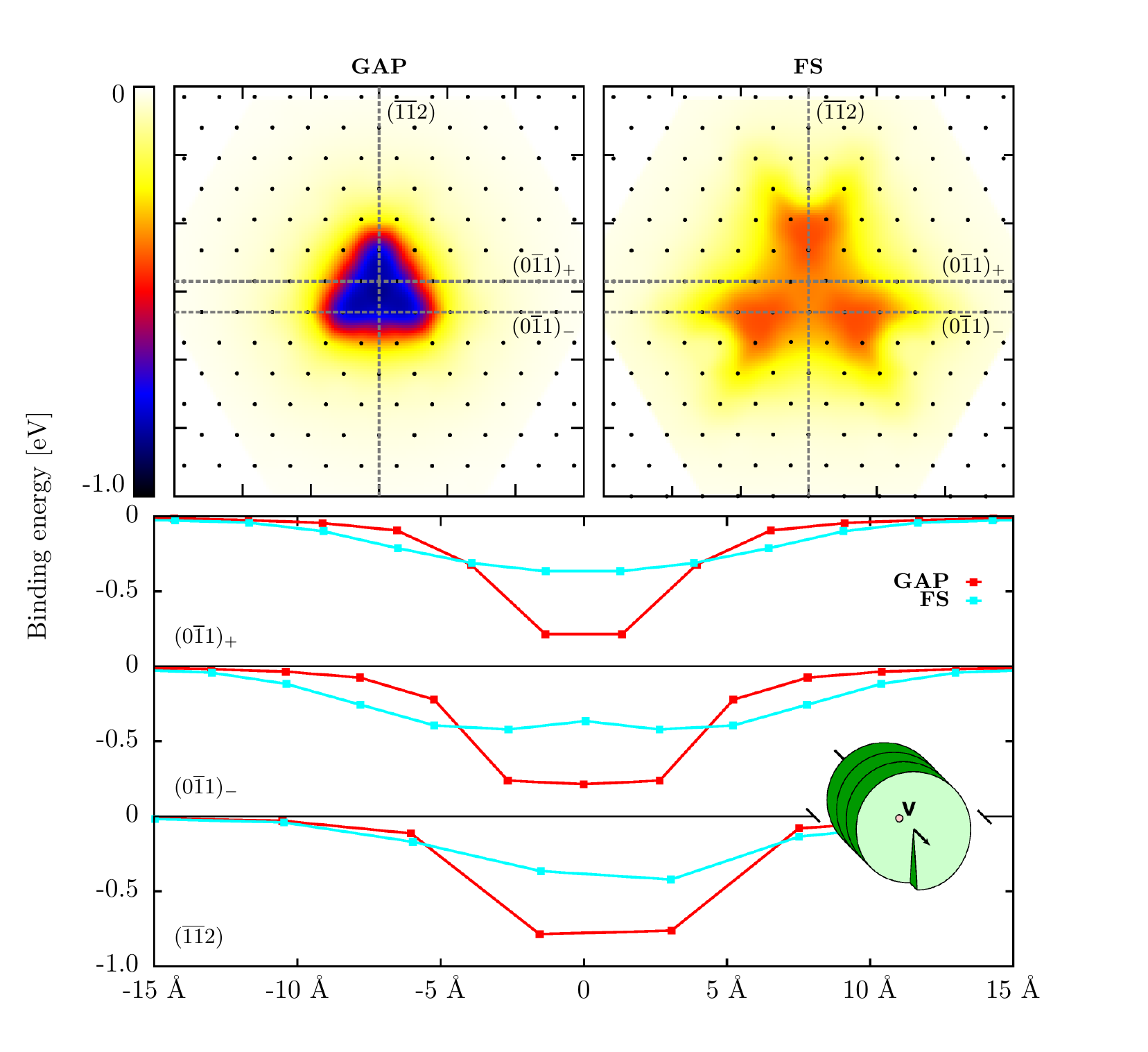}}
\normalsize{}
\vspace{-0.5cm}
\caption{Dislocation-vacancy binding energy evaluated using GAP and FS potentials. The top panels show the interpolated binding energy using a heat map, the graphs below are slices of the same along the dotted lines shown in the top panels.}
\label{figure:vacancy-map}
\end{figure}

Due to the intrinsic smoothness of the potential, it can be expected to perform well for configurations which 
contain multiple defect structures as long as the local deformation around each defect with respect to the 
corresponding configurations in the database is small. So we finally turn to  an example of the kinds of atomistic 
properties that are needed to make the connection to materials modelling on higher length scales, but are inaccessible to 
direct DFT calculations due to system size limitations imposed by the associated computational cost. Figure~\ref{figure:vacancy-map} shows the energy of a vacancy in the vicinity of a screw dislocation calculated in a 
system of over 100,000 atoms using cylindrical fixed boundary conditions   230~\AA~away from the core and 
with periodic boundary conditions applied along the dislocation line with a periodicity corresponding to three 
Burgers vectors. The Finnis-Sinclair potential  underestimates  this  interaction by a factor of two.

Although the potential developed in this work does not yet constitute a comprehensive description of tungsten under all conditions, 
we have shown  that the strategy of building a database of representative small unit cell configurations is viable, and 
will be continued with the incorporation of other crystal phases, edge dislocations, interstitials, 
etc. In addition to developing ever-more comprehensive databases and computing specific atomic scale 
properties with  first principles accuracy on which higher length scale models can be built, our long term goal is 
to discover whether,  in the context of a given material, an all-encompassing database could be assembled that 
contains a sufficient variety of neighbour environments  to be valid for any configuration encountered under 
conditions of physically realistic temperatures and pressures. If that turns out to be possible, it would herald a 
truly new era of precision for atomistic simulations in materials science.

\begin{acknowledgments}
The authors are indebted to A. De Vita and N. Bernstein for comments on the manuscript. APB is supported by a Leverhulme Early Career Fellowship and the Isaac Newton Trust. GC acknowledges support from the EPSRC grants EP/J010847/1
and EP/L014742/1. All software and data necessary for the reproduction of the results in this paper are available at www.libatoms.org.
\end{acknowledgments}

\bibliography{gap_w_1}

\begin{thebibliography}{50}%
\makeatletter
\providecommand \@ifxundefined [1]{%
 \@ifx{#1\undefined}
}%
\providecommand \@ifnum [1]{%
 \ifnum #1\expandafter \@firstoftwo
 \else \expandafter \@secondoftwo
 \fi
}%
\providecommand \@ifx [1]{%
 \ifx #1\expandafter \@firstoftwo
 \else \expandafter \@secondoftwo
 \fi
}%
\providecommand \natexlab [1]{#1}%
\providecommand \enquote  [1]{``#1''}%
\providecommand \bibnamefont  [1]{#1}%
\providecommand \bibfnamefont [1]{#1}%
\providecommand \citenamefont [1]{#1}%
\providecommand \href@noop [0]{\@secondoftwo}%
\providecommand \href [0]{\begingroup \@sanitize@url \@href}%
\providecommand \@href[1]{\@@startlink{#1}\@@href}%
\providecommand \@@href[1]{\endgroup#1\@@endlink}%
\providecommand \@sanitize@url [0]{\catcode `\\12\catcode `\$12\catcode
  `\&12\catcode `\#12\catcode `\^12\catcode `\_12\catcode `\%12\relax}%
\providecommand \@@startlink[1]{}%
\providecommand \@@endlink[0]{}%
\providecommand \url  [0]{\begingroup\@sanitize@url \@url }%
\providecommand \@url [1]{\endgroup\@href {#1}{\urlprefix }}%
\providecommand \urlprefix  [0]{URL }%
\providecommand \Eprint [0]{\href }%
\providecommand \doibase [0]{http://dx.doi.org/}%
\providecommand \selectlanguage [0]{\@gobble}%
\providecommand \bibinfo  [0]{\@secondoftwo}%
\providecommand \bibfield  [0]{\@secondoftwo}%
\providecommand \translation [1]{[#1]}%
\providecommand \BibitemOpen [0]{}%
\providecommand \bibitemStop [0]{}%
\providecommand \bibitemNoStop [0]{.\EOS\space}%
\providecommand \EOS [0]{\spacefactor3000\relax}%
\providecommand \BibitemShut  [1]{\csname bibitem#1\endcsname}%
\let\auto@bib@innerbib\@empty
\bibitem [{\citenamefont {Finnis}\ and\ \citenamefont
  {Sinclair}(1984)}]{doi:10.1080/01418618408244210}%
  \BibitemOpen
  \bibfield  {author} {\bibinfo {author} {\bibfnamefont {M.~W.}\ \bibnamefont
  {Finnis}}\ and\ \bibinfo {author} {\bibfnamefont {J.~E.}\ \bibnamefont
  {Sinclair}},\ }\href {\doibase 10.1080/01418618408244210} {\bibfield
  {journal} {\bibinfo  {journal} {Philos. Mag. A}\ }\textbf {\bibinfo {volume}
  {50}},\ \bibinfo {pages} {45} (\bibinfo {year} {1984})}\BibitemShut {NoStop}%
\bibitem [{\citenamefont {Daw}\ and\ \citenamefont
  {Baskes}(1984)}]{PhysRevB.29.6443}%
  \BibitemOpen
  \bibfield  {author} {\bibinfo {author} {\bibfnamefont {M.~S.}\ \bibnamefont
  {Daw}}\ and\ \bibinfo {author} {\bibfnamefont {M.~I.}\ \bibnamefont
  {Baskes}},\ }\href {\doibase 10.1103/PhysRevB.29.6443} {\bibfield  {journal}
  {\bibinfo  {journal} {Phys. Rev. B}\ }\textbf {\bibinfo {volume} {29}},\
  \bibinfo {pages} {6443} (\bibinfo {year} {1984})}\BibitemShut {NoStop}%
\bibitem [{\citenamefont {Ackland}\ and\ \citenamefont
  {Thetford}(1987)}]{doi:10.1080/01418618708204464}%
  \BibitemOpen
  \bibfield  {author} {\bibinfo {author} {\bibfnamefont {G.~J.}\ \bibnamefont
  {Ackland}}\ and\ \bibinfo {author} {\bibfnamefont {R.}~\bibnamefont
  {Thetford}},\ }\href {\doibase 10.1080/01418618708204464} {\bibfield
  {journal} {\bibinfo  {journal} {Philos. Mag. A}\ }\textbf {\bibinfo {volume}
  {56}},\ \bibinfo {pages} {15} (\bibinfo {year} {1987})}\BibitemShut {NoStop}%
\bibitem [{\citenamefont {Sutton}\ and\ \citenamefont
  {Chen}(1990)}]{doi:10.1080/09500839008206493}%
  \BibitemOpen
  \bibfield  {author} {\bibinfo {author} {\bibfnamefont {A.~P.}\ \bibnamefont
  {Sutton}}\ and\ \bibinfo {author} {\bibfnamefont {J.}~\bibnamefont {Chen}},\
  }\href {\doibase 10.1080/09500839008206493} {\bibfield  {journal} {\bibinfo
  {journal} {Philos. Mag. Lett.}\ }\textbf {\bibinfo {volume} {61}},\ \bibinfo
  {pages} {139} (\bibinfo {year} {1990})}\BibitemShut {NoStop}%
\bibitem [{\citenamefont {Wang}\ \emph {et~al.}(2014)\citenamefont {Wang},
  \citenamefont {Zhou}, \citenamefont {Li},\ and\ \citenamefont
  {Hou}}]{0965-0393-22-1-015004}%
  \BibitemOpen
  \bibfield  {author} {\bibinfo {author} {\bibfnamefont {J.}~\bibnamefont
  {Wang}}, \bibinfo {author} {\bibfnamefont {Y.~L.}\ \bibnamefont {Zhou}},
  \bibinfo {author} {\bibfnamefont {M.}~\bibnamefont {Li}}, \ and\ \bibinfo
  {author} {\bibfnamefont {Q.}~\bibnamefont {Hou}},\ }\href {\doibase
  10.1088/0965-0393/22/1/015004} {\bibfield  {journal} {\bibinfo  {journal}
  {Modell. Simul. Mater. Sci. Eng.}\ }\textbf {\bibinfo {volume} {22}},\
  \bibinfo {pages} {015004} (\bibinfo {year} {2014})}\BibitemShut {NoStop}%
\bibitem [{\citenamefont {Ercolessi}\ and\ \citenamefont
  {Adams}(1994)}]{0295-5075-26-8-005}%
  \BibitemOpen
  \bibfield  {author} {\bibinfo {author} {\bibfnamefont {F.}~\bibnamefont
  {Ercolessi}}\ and\ \bibinfo {author} {\bibfnamefont {J.~B.}\ \bibnamefont
  {Adams}},\ }\href {\doibase 10.1209/0295-5075/26/8/005} {\bibfield  {journal}
  {\bibinfo  {journal} {Europhys. Lett.}\ }\textbf {\bibinfo {volume} {26}},\
  \bibinfo {pages} {583} (\bibinfo {year} {1994})}\BibitemShut {NoStop}%
\bibitem [{\citenamefont {Baskes}(1992)}]{PhysRevB.46.2727}%
  \BibitemOpen
  \bibfield  {author} {\bibinfo {author} {\bibfnamefont {M.~I.}\ \bibnamefont
  {Baskes}},\ }\href {\doibase 10.1103/PhysRevB.46.2727} {\bibfield  {journal}
  {\bibinfo  {journal} {Phys. Rev. B}\ }\textbf {\bibinfo {volume} {46}},\
  \bibinfo {pages} {2727} (\bibinfo {year} {1992})}\BibitemShut {NoStop}%
\bibitem [{\citenamefont {Wang}\ and\ \citenamefont
  {Boercker}(1995)}]{:/content/aip/journal/jap/78/1/10.1063/1.360661}%
  \BibitemOpen
  \bibfield  {author} {\bibinfo {author} {\bibfnamefont {Y.~R.}\ \bibnamefont
  {Wang}}\ and\ \bibinfo {author} {\bibfnamefont {D.~B.}\ \bibnamefont
  {Boercker}},\ }\href {\doibase 10.1063/1.360661} {\bibfield  {journal}
  {\bibinfo  {journal} {J. Appl. Phys.}\ }\textbf {\bibinfo {volume} {78}},\
  \bibinfo {pages} {122} (\bibinfo {year} {1995})}\BibitemShut {NoStop}%
\bibitem [{\citenamefont {Lee}\ \emph {et~al.}(2001)\citenamefont {Lee},
  \citenamefont {Baskes}, \citenamefont {Kim},\ and\ \citenamefont
  {Koo~Cho}}]{PhysRevB.64.184102}%
  \BibitemOpen
  \bibfield  {author} {\bibinfo {author} {\bibfnamefont {B.-J.}\ \bibnamefont
  {Lee}}, \bibinfo {author} {\bibfnamefont {M.~I.}\ \bibnamefont {Baskes}},
  \bibinfo {author} {\bibfnamefont {H.}~\bibnamefont {Kim}}, \ and\ \bibinfo
  {author} {\bibfnamefont {Y.}~\bibnamefont {Koo~Cho}},\ }\href {\doibase
  10.1103/PhysRevB.64.184102} {\bibfield  {journal} {\bibinfo  {journal} {Phys.
  Rev. B}\ }\textbf {\bibinfo {volume} {64}},\ \bibinfo {pages} {184102}
  (\bibinfo {year} {2001})}\BibitemShut {NoStop}%
\bibitem [{\citenamefont {Marinica}\ \emph {et~al.}(2013)\citenamefont
  {Marinica}, \citenamefont {Ventelon}, \citenamefont {Gilbert}, \citenamefont
  {Proville}, \citenamefont {Dudarev}, \citenamefont {Marian}, \citenamefont
  {Bencteux},\ and\ \citenamefont {Willaime}}]{0953-8984-25-39-395502}%
  \BibitemOpen
  \bibfield  {author} {\bibinfo {author} {\bibfnamefont {M.-C.}\ \bibnamefont
  {Marinica}}, \bibinfo {author} {\bibfnamefont {L.}~\bibnamefont {Ventelon}},
  \bibinfo {author} {\bibfnamefont {M.~R.}\ \bibnamefont {Gilbert}}, \bibinfo
  {author} {\bibfnamefont {L.}~\bibnamefont {Proville}}, \bibinfo {author}
  {\bibfnamefont {S.~L.}\ \bibnamefont {Dudarev}}, \bibinfo {author}
  {\bibfnamefont {J.}~\bibnamefont {Marian}}, \bibinfo {author} {\bibfnamefont
  {G.}~\bibnamefont {Bencteux}}, \ and\ \bibinfo {author} {\bibfnamefont
  {F.}~\bibnamefont {Willaime}},\ }\href {\doibase
  10.1088/0953-8984/25/39/395502} {\bibfield  {journal} {\bibinfo  {journal}
  {J. Phys.: Condens. Matter}\ }\textbf {\bibinfo {volume} {25}},\ \bibinfo
  {pages} {395502} (\bibinfo {year} {2013})}\BibitemShut {NoStop}%
\bibitem [{\citenamefont {Mrovec}\ \emph {et~al.}(2007)\citenamefont {Mrovec},
  \citenamefont {Gr\"oger}, \citenamefont {Bailey}, \citenamefont
  {Nguyen-Manh}, \citenamefont {Els\"asser},\ and\ \citenamefont
  {Vitek}}]{PhysRevB.75.104119}%
  \BibitemOpen
  \bibfield  {author} {\bibinfo {author} {\bibfnamefont {M.}~\bibnamefont
  {Mrovec}}, \bibinfo {author} {\bibfnamefont {R.}~\bibnamefont {Gr\"oger}},
  \bibinfo {author} {\bibfnamefont {A.~G.}\ \bibnamefont {Bailey}}, \bibinfo
  {author} {\bibfnamefont {D.}~\bibnamefont {Nguyen-Manh}}, \bibinfo {author}
  {\bibfnamefont {C.}~\bibnamefont {Els\"asser}}, \ and\ \bibinfo {author}
  {\bibfnamefont {V.}~\bibnamefont {Vitek}},\ }\href {\doibase
  10.1103/PhysRevB.75.104119} {\bibfield  {journal} {\bibinfo  {journal} {Phys.
  Rev. B}\ }\textbf {\bibinfo {volume} {75}},\ \bibinfo {pages} {104119}
  (\bibinfo {year} {2007})}\BibitemShut {NoStop}%
\bibitem [{\citenamefont {Ahlgren}\ \emph {et~al.}(2010)\citenamefont
  {Ahlgren}, \citenamefont {Heinola}, \citenamefont {Juslin},\ and\
  \citenamefont {Kuronen}}]{:/content/aip/journal/jap/107/3/10.1063/1.3298466}%
  \BibitemOpen
  \bibfield  {author} {\bibinfo {author} {\bibfnamefont {T.}~\bibnamefont
  {Ahlgren}}, \bibinfo {author} {\bibfnamefont {K.}~\bibnamefont {Heinola}},
  \bibinfo {author} {\bibfnamefont {N.}~\bibnamefont {Juslin}}, \ and\ \bibinfo
  {author} {\bibfnamefont {A.}~\bibnamefont {Kuronen}},\ }\href {\doibase
  10.1063/1.3298466} {\bibfield  {journal} {\bibinfo  {journal} {J. Appl.
  Phys.}\ }\textbf {\bibinfo {volume} {107}},\ \bibinfo {eid} {033516}
  (\bibinfo {year} {2010})}\BibitemShut {NoStop}%
\bibitem [{\citenamefont {Li}\ \emph {et~al.}(2011)\citenamefont {Li},
  \citenamefont {Shu}, \citenamefont {Liu}, \citenamefont {Gao},\ and\
  \citenamefont {Lu}}]{Li201112}%
  \BibitemOpen
  \bibfield  {author} {\bibinfo {author} {\bibfnamefont {X.-C.}\ \bibnamefont
  {Li}}, \bibinfo {author} {\bibfnamefont {X.}~\bibnamefont {Shu}}, \bibinfo
  {author} {\bibfnamefont {Y.-N.}\ \bibnamefont {Liu}}, \bibinfo {author}
  {\bibfnamefont {F.}~\bibnamefont {Gao}}, \ and\ \bibinfo {author}
  {\bibfnamefont {G.-H.}\ \bibnamefont {Lu}},\ }\href {\doibase
  10.1016/j.jnucmat.2010.10.020} {\bibfield  {journal} {\bibinfo  {journal} {J.
  Nucl. Mater.}\ }\textbf {\bibinfo {volume} {408}},\ \bibinfo {pages} {12 }
  (\bibinfo {year} {2011})}\BibitemShut {NoStop}%
\bibitem [{\citenamefont {Moriarty}(1988)}]{PhysRevB.38.3199}%
  \BibitemOpen
  \bibfield  {author} {\bibinfo {author} {\bibfnamefont {J.~A.}\ \bibnamefont
  {Moriarty}},\ }\href {\doibase 10.1103/PhysRevB.38.3199} {\bibfield
  {journal} {\bibinfo  {journal} {Phys. Rev. B}\ }\textbf {\bibinfo {volume}
  {38}},\ \bibinfo {pages} {3199} (\bibinfo {year} {1988})}\BibitemShut
  {NoStop}%
\bibitem [{\citenamefont {Xu}\ and\ \citenamefont
  {Moriarty}(1996)}]{PhysRevB.54.6941}%
  \BibitemOpen
  \bibfield  {author} {\bibinfo {author} {\bibfnamefont {W.}~\bibnamefont
  {Xu}}\ and\ \bibinfo {author} {\bibfnamefont {J.~A.}\ \bibnamefont
  {Moriarty}},\ }\href {\doibase 10.1103/PhysRevB.54.6941} {\bibfield
  {journal} {\bibinfo  {journal} {Phys. Rev. B}\ }\textbf {\bibinfo {volume}
  {54}},\ \bibinfo {pages} {6941} (\bibinfo {year} {1996})}\BibitemShut
  {NoStop}%
\bibitem [{\citenamefont {Mrovec}\ \emph {et~al.}(2004)\citenamefont {Mrovec},
  \citenamefont {Nguyen-Manh}, \citenamefont {Pettifor},\ and\ \citenamefont
  {Vitek}}]{PhysRevB.69.094115}%
  \BibitemOpen
  \bibfield  {author} {\bibinfo {author} {\bibfnamefont {M.}~\bibnamefont
  {Mrovec}}, \bibinfo {author} {\bibfnamefont {D.}~\bibnamefont {Nguyen-Manh}},
  \bibinfo {author} {\bibfnamefont {D.~G.}\ \bibnamefont {Pettifor}}, \ and\
  \bibinfo {author} {\bibfnamefont {V.}~\bibnamefont {Vitek}},\ }\href
  {\doibase 10.1103/PhysRevB.69.094115} {\bibfield  {journal} {\bibinfo
  {journal} {Phys. Rev. B}\ }\textbf {\bibinfo {volume} {69}},\ \bibinfo
  {pages} {094115} (\bibinfo {year} {2004})}\BibitemShut {NoStop}%
\bibitem [{\citenamefont {Matthews}\ \emph {et~al.}(2007)\citenamefont
  {Matthews}, \citenamefont {Edwards}, \citenamefont {Hirai}, \citenamefont
  {Kear}, \citenamefont {Lioure}, \citenamefont {Lomas}, \citenamefont
  {Loving}, \citenamefont {Lungu}, \citenamefont {Maier}, \citenamefont
  {Mertens} \emph {et~al.}}]{1402-4896-2007-T128-027}%
  \BibitemOpen
  \bibfield  {author} {\bibinfo {author} {\bibfnamefont {G.~F.}\ \bibnamefont
  {Matthews}}, \bibinfo {author} {\bibfnamefont {P.}~\bibnamefont {Edwards}},
  \bibinfo {author} {\bibfnamefont {T.}~\bibnamefont {Hirai}}, \bibinfo
  {author} {\bibfnamefont {M.}~\bibnamefont {Kear}}, \bibinfo {author}
  {\bibfnamefont {A.}~\bibnamefont {Lioure}}, \bibinfo {author} {\bibfnamefont
  {P.}~\bibnamefont {Lomas}}, \bibinfo {author} {\bibfnamefont
  {A.}~\bibnamefont {Loving}}, \bibinfo {author} {\bibfnamefont
  {C.}~\bibnamefont {Lungu}}, \bibinfo {author} {\bibfnamefont
  {H.}~\bibnamefont {Maier}}, \bibinfo {author} {\bibfnamefont
  {P.}~\bibnamefont {Mertens}},  \emph {et~al.},\ }\href {\doibase
  10.1088/0031-8949/2007/T128/027} {\bibfield  {journal} {\bibinfo  {journal}
  {Phys. Scripta}\ }\textbf {\bibinfo {volume} {2007}},\ \bibinfo {pages} {137}
  (\bibinfo {year} {2007})}\BibitemShut {NoStop}%
\bibitem [{\citenamefont {Neu}\ \emph {et~al.}(2007)\citenamefont {Neu},
  \citenamefont {Balden}, \citenamefont {Bobkov}, \citenamefont {Dux},
  \citenamefont {Gruber}, \citenamefont {Herrmann}, \citenamefont {Kallenbach},
  \citenamefont {Kaufmann}, \citenamefont {Maggi}, \citenamefont {Maier} \emph
  {et~al.}}]{0741-3335-49-12B-S04}%
  \BibitemOpen
  \bibfield  {author} {\bibinfo {author} {\bibfnamefont {R.}~\bibnamefont
  {Neu}}, \bibinfo {author} {\bibfnamefont {M.}~\bibnamefont {Balden}},
  \bibinfo {author} {\bibfnamefont {V.}~\bibnamefont {Bobkov}}, \bibinfo
  {author} {\bibfnamefont {R.}~\bibnamefont {Dux}}, \bibinfo {author}
  {\bibfnamefont {O.}~\bibnamefont {Gruber}}, \bibinfo {author} {\bibfnamefont
  {A.}~\bibnamefont {Herrmann}}, \bibinfo {author} {\bibfnamefont
  {A.}~\bibnamefont {Kallenbach}}, \bibinfo {author} {\bibfnamefont
  {M.}~\bibnamefont {Kaufmann}}, \bibinfo {author} {\bibfnamefont {C.~F.}\
  \bibnamefont {Maggi}}, \bibinfo {author} {\bibfnamefont {H.}~\bibnamefont
  {Maier}},  \emph {et~al.},\ }\href {\doibase 10.1088/0741-3335/49/12B/S04}
  {\bibfield  {journal} {\bibinfo  {journal} {Plasma Phys. Controlled Fusion}\
  }\textbf {\bibinfo {volume} {49}},\ \bibinfo {pages} {B59} (\bibinfo {year}
  {2007})}\BibitemShut {NoStop}%
\bibitem [{\citenamefont {Pitts}\ \emph {et~al.}(2013)\citenamefont {Pitts},
  \citenamefont {Carpentier}, \citenamefont {Escourbiac}, \citenamefont
  {Hirai}, \citenamefont {Komarov}, \citenamefont {Lisgo}, \citenamefont
  {Kukushkin}, \citenamefont {Loarte}, \citenamefont {Merola}, \citenamefont
  {Sashala} \emph {et~al.}}]{Pitts2013S48}%
  \BibitemOpen
  \bibfield  {author} {\bibinfo {author} {\bibfnamefont {R.}~\bibnamefont
  {Pitts}}, \bibinfo {author} {\bibfnamefont {S.}~\bibnamefont {Carpentier}},
  \bibinfo {author} {\bibfnamefont {F.}~\bibnamefont {Escourbiac}}, \bibinfo
  {author} {\bibfnamefont {T.}~\bibnamefont {Hirai}}, \bibinfo {author}
  {\bibfnamefont {V.}~\bibnamefont {Komarov}}, \bibinfo {author} {\bibfnamefont
  {S.}~\bibnamefont {Lisgo}}, \bibinfo {author} {\bibfnamefont
  {A.}~\bibnamefont {Kukushkin}}, \bibinfo {author} {\bibfnamefont
  {A.}~\bibnamefont {Loarte}}, \bibinfo {author} {\bibfnamefont
  {M.}~\bibnamefont {Merola}}, \bibinfo {author} {\bibfnamefont
  {A.}~\bibnamefont {Sashala}},  \emph {et~al.},\ }\href {\doibase
  10.1016/j.jnucmat.2013.01.008} {\bibfield  {journal} {\bibinfo  {journal} {J.
  Nucl. Mater.}\ }\textbf {\bibinfo {volume} {438, Supplement}},\ \bibinfo
  {pages} {S48 } (\bibinfo {year} {2013})}\BibitemShut {NoStop}%
\bibitem [{\citenamefont {Bernstein}\ \emph {et~al.}(2009)\citenamefont
  {Bernstein}, \citenamefont {Kermode},\ and\ \citenamefont
  {Cs\'{a}nyi}}]{Bernstein2009}%
  \BibitemOpen
  \bibfield  {author} {\bibinfo {author} {\bibfnamefont {N.}~\bibnamefont
  {Bernstein}}, \bibinfo {author} {\bibfnamefont {J.~R.}\ \bibnamefont
  {Kermode}}, \ and\ \bibinfo {author} {\bibfnamefont {G.}~\bibnamefont
  {Cs\'{a}nyi}},\ }\href {\doibase 10.1088/0034-4885/72/2/026501} {\bibfield
  {journal} {\bibinfo  {journal} {Rep. Prog. Phys.}\ }\textbf {\bibinfo
  {volume} {72}},\ \bibinfo {pages} {026501} (\bibinfo {year}
  {2009})}\BibitemShut {NoStop}%
\bibitem [{\citenamefont {Vita}\ and\ \citenamefont {Car}(1997)}]{Devita1997}%
  \BibitemOpen
  \bibfield  {author} {\bibinfo {author} {\bibfnamefont {A.~D.}\ \bibnamefont
  {Vita}}\ and\ \bibinfo {author} {\bibfnamefont {R.}~\bibnamefont {Car}},\
  }\href {\doibase 10.1557/PROC-491-473} {\bibfield  {journal} {\bibinfo
  {journal} {MRS Bull.}\ }\textbf {\bibinfo {volume} {491}},\ \bibinfo {pages}
  {473} (\bibinfo {year} {1997})}\BibitemShut {NoStop}%
\bibitem [{\citenamefont {Behler}\ and\ \citenamefont
  {Parrinello}(2007)}]{PhysRevLett.98.146401}%
  \BibitemOpen
  \bibfield  {author} {\bibinfo {author} {\bibfnamefont {J.}~\bibnamefont
  {Behler}}\ and\ \bibinfo {author} {\bibfnamefont {M.}~\bibnamefont
  {Parrinello}},\ }\href {\doibase 10.1103/PhysRevLett.98.146401} {\bibfield
  {journal} {\bibinfo  {journal} {Phys. Rev. Lett.}\ }\textbf {\bibinfo
  {volume} {98}},\ \bibinfo {pages} {146401} (\bibinfo {year}
  {2007})}\BibitemShut {NoStop}%
\bibitem [{\citenamefont {Behler}\ \emph {et~al.}(2008)\citenamefont {Behler},
  \citenamefont {Marto\v{n}\'{a}k}, \citenamefont {Donadio},\ and\
  \citenamefont {Parrinello}}]{PhysRevLett.100.185501}%
  \BibitemOpen
  \bibfield  {author} {\bibinfo {author} {\bibfnamefont {J.}~\bibnamefont
  {Behler}}, \bibinfo {author} {\bibfnamefont {R.}~\bibnamefont
  {Marto\v{n}\'{a}k}}, \bibinfo {author} {\bibfnamefont {D.}~\bibnamefont
  {Donadio}}, \ and\ \bibinfo {author} {\bibfnamefont {M.}~\bibnamefont
  {Parrinello}},\ }\href {\doibase 10.1103/PhysRevLett.100.185501} {\bibfield
  {journal} {\bibinfo  {journal} {Phys. Rev. Lett.}\ }\textbf {\bibinfo
  {volume} {100}},\ \bibinfo {pages} {185501} (\bibinfo {year}
  {2008})}\BibitemShut {NoStop}%
\bibitem [{\citenamefont {Artrith}\ and\ \citenamefont
  {Behler}(2012)}]{PhysRevB.85.045439}%
  \BibitemOpen
  \bibfield  {author} {\bibinfo {author} {\bibfnamefont {N.}~\bibnamefont
  {Artrith}}\ and\ \bibinfo {author} {\bibfnamefont {J.}~\bibnamefont
  {Behler}},\ }\href {\doibase 10.1103/PhysRevB.85.045439} {\bibfield
  {journal} {\bibinfo  {journal} {Phys. Rev. B}\ }\textbf {\bibinfo {volume}
  {85}},\ \bibinfo {pages} {045439} (\bibinfo {year} {2012})}\BibitemShut
  {NoStop}%
\bibitem [{\citenamefont {Ischtwan}\ and\ \citenamefont
  {Collins}(1994)}]{:/content/aip/journal/jcp/100/11/10.1063/1.466801}%
  \BibitemOpen
  \bibfield  {author} {\bibinfo {author} {\bibfnamefont {J.}~\bibnamefont
  {Ischtwan}}\ and\ \bibinfo {author} {\bibfnamefont {M.~A.}\ \bibnamefont
  {Collins}},\ }\href {\doibase 10.1063/1.466801} {\bibfield  {journal}
  {\bibinfo  {journal} {J. Chem. Phys.}\ }\textbf {\bibinfo {volume} {100}},\
  \bibinfo {pages} {8080} (\bibinfo {year} {1994})}\BibitemShut {NoStop}%
\bibitem [{\citenamefont {Collins}(2002)}]{Collins2002}%
  \BibitemOpen
  \bibfield  {author} {\bibinfo {author} {\bibfnamefont {M.~A.}\ \bibnamefont
  {Collins}},\ }\href {\doibase 10.1007/s00214-002-0383-5} {\bibfield
  {journal} {\bibinfo  {journal} {Theor. Chem. Acc.}\ }\textbf {\bibinfo
  {volume} {108}},\ \bibinfo {pages} {313} (\bibinfo {year}
  {2002})}\BibitemShut {NoStop}%
\bibitem [{\citenamefont {Zhang}\ \emph {et~al.}(2004)\citenamefont {Zhang},
  \citenamefont {Zou}, \citenamefont {Harding},\ and\ \citenamefont
  {Bowman}}]{doi:10.1021/jp048339l}%
  \BibitemOpen
  \bibfield  {author} {\bibinfo {author} {\bibfnamefont {X.}~\bibnamefont
  {Zhang}}, \bibinfo {author} {\bibfnamefont {S.}~\bibnamefont {Zou}}, \bibinfo
  {author} {\bibfnamefont {L.~B.}\ \bibnamefont {Harding}}, \ and\ \bibinfo
  {author} {\bibfnamefont {J.~M.}\ \bibnamefont {Bowman}},\ }\href {\doibase
  10.1021/jp048339l} {\bibfield  {journal} {\bibinfo  {journal} {J. Phys. Chem.
  A}\ }\textbf {\bibinfo {volume} {108}},\ \bibinfo {pages} {8980} (\bibinfo
  {year} {2004})}\BibitemShut {NoStop}%
\bibitem [{\citenamefont {Huang}\ \emph {et~al.}(2005)\citenamefont {Huang},
  \citenamefont {Braams},\ and\ \citenamefont
  {Bowman}}]{:/content/aip/journal/jcp/122/4/10.1063/1.1834500}%
  \BibitemOpen
  \bibfield  {author} {\bibinfo {author} {\bibfnamefont {X.}~\bibnamefont
  {Huang}}, \bibinfo {author} {\bibfnamefont {B.~J.}\ \bibnamefont {Braams}}, \
  and\ \bibinfo {author} {\bibfnamefont {J.~M.}\ \bibnamefont {Bowman}},\
  }\href {\doibase 10.1063/1.1834500} {\bibfield  {journal} {\bibinfo
  {journal} {J. Chem. Phys.}\ }\textbf {\bibinfo {volume} {122}},\ \bibinfo
  {eid} {044308} (\bibinfo {year} {2005})}\BibitemShut {NoStop}%
\bibitem [{\citenamefont {Xie}\ \emph {et~al.}(2005)\citenamefont {Xie},
  \citenamefont {Braams},\ and\ \citenamefont
  {Bowman}}]{:/content/aip/journal/jcp/122/22/10.1063/1.1927529}%
  \BibitemOpen
  \bibfield  {author} {\bibinfo {author} {\bibfnamefont {Z.}~\bibnamefont
  {Xie}}, \bibinfo {author} {\bibfnamefont {B.~J.}\ \bibnamefont {Braams}}, \
  and\ \bibinfo {author} {\bibfnamefont {J.~M.}\ \bibnamefont {Bowman}},\
  }\href {\doibase 10.1063/1.1927529} {\bibfield  {journal} {\bibinfo
  {journal} {J. Chem. Phys.}\ }\textbf {\bibinfo {volume} {122}},\ \bibinfo
  {eid} {224307} (\bibinfo {year} {2005})}\BibitemShut {NoStop}%
\bibitem [{\citenamefont {Bart\'{o}k}\ \emph {et~al.}(2010)\citenamefont
  {Bart\'{o}k}, \citenamefont {Payne}, \citenamefont {Kondor},\ and\
  \citenamefont {Cs\'{a}nyi}}]{PhysRevLett.104.136403}%
  \BibitemOpen
  \bibfield  {author} {\bibinfo {author} {\bibfnamefont {A.~P.}\ \bibnamefont
  {Bart\'{o}k}}, \bibinfo {author} {\bibfnamefont {M.~C.}\ \bibnamefont
  {Payne}}, \bibinfo {author} {\bibfnamefont {R.}~\bibnamefont {Kondor}}, \
  and\ \bibinfo {author} {\bibfnamefont {G.}~\bibnamefont {Cs\'{a}nyi}},\
  }\href {\doibase 10.1103/PhysRevLett.104.136403} {\bibfield  {journal}
  {\bibinfo  {journal} {Phys. Rev. Lett.}\ }\textbf {\bibinfo {volume} {104}},\
  \bibinfo {pages} {136403} (\bibinfo {year} {2010})}\BibitemShut {NoStop}%
\bibitem [{\citenamefont {Bart\'{o}k}\ \emph {et~al.}(2013)\citenamefont
  {Bart\'{o}k}, \citenamefont {Kondor},\ and\ \citenamefont
  {Cs\'{a}nyi}}]{PhysRevB.87.184115}%
  \BibitemOpen
  \bibfield  {author} {\bibinfo {author} {\bibfnamefont {A.~P.}\ \bibnamefont
  {Bart\'{o}k}}, \bibinfo {author} {\bibfnamefont {R.}~\bibnamefont {Kondor}},
  \ and\ \bibinfo {author} {\bibfnamefont {G.}~\bibnamefont {Cs\'{a}nyi}},\
  }\href {\doibase 10.1103/PhysRevB.87.184115} {\bibfield  {journal} {\bibinfo
  {journal} {Phys. Rev. B}\ }\textbf {\bibinfo {volume} {87}},\ \bibinfo
  {pages} {184115} (\bibinfo {year} {2013})}\BibitemShut {NoStop}%
\bibitem [{\citenamefont {Bart\'ok}\ \emph {et~al.}(2013)\citenamefont
  {Bart\'ok}, \citenamefont {Gillan}, \citenamefont {Manby},\ and\
  \citenamefont {Cs\'anyi}}]{PhysRevB.88.054104}%
  \BibitemOpen
  \bibfield  {author} {\bibinfo {author} {\bibfnamefont {A.~P.}\ \bibnamefont
  {Bart\'ok}}, \bibinfo {author} {\bibfnamefont {M.~J.}\ \bibnamefont
  {Gillan}}, \bibinfo {author} {\bibfnamefont {F.~R.}\ \bibnamefont {Manby}}, \
  and\ \bibinfo {author} {\bibfnamefont {G.}~\bibnamefont {Cs\'anyi}},\ }\href
  {\doibase 10.1103/PhysRevB.88.054104} {\bibfield  {journal} {\bibinfo
  {journal} {Phys. Rev. B}\ }\textbf {\bibinfo {volume} {88}},\ \bibinfo
  {pages} {054104} (\bibinfo {year} {2013})}\BibitemShut {NoStop}%
\bibitem [{\citenamefont {Gillan}\ \emph {et~al.}(2013)\citenamefont {Gillan},
  \citenamefont {Alf\`{e}}, \citenamefont {Bart\'{o}k},\ and\ \citenamefont
  {Cs\'{a}nyi}}]{:/content/aip/journal/jcp/139/24/10.1063/1.4852182}%
  \BibitemOpen
  \bibfield  {author} {\bibinfo {author} {\bibfnamefont {M.~J.}\ \bibnamefont
  {Gillan}}, \bibinfo {author} {\bibfnamefont {D.}~\bibnamefont {Alf\`{e}}},
  \bibinfo {author} {\bibfnamefont {A.~P.}\ \bibnamefont {Bart\'{o}k}}, \ and\
  \bibinfo {author} {\bibfnamefont {G.}~\bibnamefont {Cs\'{a}nyi}},\ }\href
  {\doibase http://dx.doi.org/10.1063/1.4852182} {\bibfield  {journal}
  {\bibinfo  {journal} {J. Chem. Phys.}\ }\textbf {\bibinfo {volume} {139}},\
  \bibinfo {eid} {244504} (\bibinfo {year} {2013})}\BibitemShut {NoStop}%
\bibitem [{\citenamefont {Rupp}\ \emph {et~al.}(2012)\citenamefont {Rupp},
  \citenamefont {Tkatchenko}, \citenamefont {M\"uller},\ and\ \citenamefont
  {von Lilienfeld}}]{PhysRevLett.108.058301}%
  \BibitemOpen
  \bibfield  {author} {\bibinfo {author} {\bibfnamefont {M.}~\bibnamefont
  {Rupp}}, \bibinfo {author} {\bibfnamefont {A.}~\bibnamefont {Tkatchenko}},
  \bibinfo {author} {\bibfnamefont {K.-R.}\ \bibnamefont {M\"uller}}, \ and\
  \bibinfo {author} {\bibfnamefont {O.~A.}\ \bibnamefont {von Lilienfeld}},\
  }\href {\doibase 10.1103/PhysRevLett.108.058301} {\bibfield  {journal}
  {\bibinfo  {journal} {Phys. Rev. Lett.}\ }\textbf {\bibinfo {volume} {108}},\
  \bibinfo {pages} {058301} (\bibinfo {year} {2012})}\BibitemShut {NoStop}%
\bibitem [{\citenamefont {MacKay}(2003)}]{mackay2003information}%
  \BibitemOpen
  \bibfield  {author} {\bibinfo {author} {\bibfnamefont {D.}~\bibnamefont
  {MacKay}},\ }\href {http://books.google.co.uk/books?id=i0XJngEACAAJ} {\emph
  {\bibinfo {title} {Information Theory, Inference and Learning Algorithms}}}\
  (\bibinfo  {publisher} {Cambridge University Press},\ \bibinfo {year}
  {2003})\BibitemShut {NoStop}%
\bibitem [{\citenamefont {Rasmussen}\ and\ \citenamefont
  {Williams}(2006)}]{rasmussen2006gaussian}%
  \BibitemOpen
  \bibfield  {author} {\bibinfo {author} {\bibfnamefont {C.}~\bibnamefont
  {Rasmussen}}\ and\ \bibinfo {author} {\bibfnamefont {C.}~\bibnamefont
  {Williams}},\ }\href {http://books.google.co.uk/books?id=vWtwQgAACAAJ} {\emph
  {\bibinfo {title} {Gaussian Processes for Machine Learning}}}\ (\bibinfo
  {publisher} {University Press Group Limited},\ \bibinfo {year}
  {2006})\BibitemShut {NoStop}%
\bibitem [{\citenamefont {Clark}\ \emph {et~al.}(2005)\citenamefont {Clark},
  \citenamefont {Segall}, \citenamefont {Pickard}, \citenamefont {Hasnip},
  \citenamefont {Probert}, \citenamefont {Refson},\ and\ \citenamefont
  {Payne}}]{clark05-zkryst}%
  \BibitemOpen
  \bibfield  {author} {\bibinfo {author} {\bibfnamefont {S.~J.}\ \bibnamefont
  {Clark}}, \bibinfo {author} {\bibfnamefont {M.~D.}\ \bibnamefont {Segall}},
  \bibinfo {author} {\bibfnamefont {C.~J.}\ \bibnamefont {Pickard}}, \bibinfo
  {author} {\bibfnamefont {P.~J.}\ \bibnamefont {Hasnip}}, \bibinfo {author}
  {\bibfnamefont {M.~J.}\ \bibnamefont {Probert}}, \bibinfo {author}
  {\bibfnamefont {K.}~\bibnamefont {Refson}}, \ and\ \bibinfo {author}
  {\bibfnamefont {M.~C.}\ \bibnamefont {Payne}},\ }\href {\doibase
  10.1524/zkri.220.5.567.65075} {\bibfield  {journal} {\bibinfo  {journal} {Z.
  Krystallogr.}\ }\textbf {\bibinfo {volume} {220}},\ \bibinfo {pages} {567}
  (\bibinfo {year} {2005})}\BibitemShut {NoStop}%
\bibitem [{\citenamefont {Snelson}\ and\ \citenamefont
  {Ghahramani}(2006)}]{NIPS2005_543}%
  \BibitemOpen
  \bibfield  {author} {\bibinfo {author} {\bibfnamefont {E.}~\bibnamefont
  {Snelson}}\ and\ \bibinfo {author} {\bibfnamefont {Z.}~\bibnamefont
  {Ghahramani}},\ }in\ \href
  {http://books.nips.cc/papers/files/nips18/NIPS2005_0543.pdf} {\emph {\bibinfo
  {booktitle} {Advances in Neural Information Processing Systems 18}}}\
  (\bibinfo  {publisher} {MIT Press},\ \bibinfo {year} {2006})\ pp.\ \bibinfo
  {pages} {1257--1264}\BibitemShut {NoStop}%
\bibitem [{\citenamefont {Bart\'{o}k}(2010)}]{2010PhDT}%
  \BibitemOpen
  \bibfield  {author} {\bibinfo {author} {\bibfnamefont {A.~P.}\ \bibnamefont
  {Bart\'{o}k}},\ }\href {http://arxiv.org/abs/1003.2817} {Ph.D. thesis},\
  \bibinfo  {school} {University of Cambridge} (\bibinfo {year}
  {2010})\BibitemShut {NoStop}%
\bibitem [{\citenamefont {Szlachta}(2013)}]{2013PhDT}%
  \BibitemOpen
  \bibfield  {author} {\bibinfo {author} {\bibfnamefont {W.~J.}\ \bibnamefont
  {Szlachta}},\ }\href {http://arxiv.org/abs/1403.3291} {Ph.D. thesis},\
  \bibinfo  {school} {University of Cambridge} (\bibinfo {year}
  {2013})\BibitemShut {NoStop}%
\bibitem [{\citenamefont {Vitek}\ and\ \citenamefont
  {Kroupa}(1969)}]{doi:10.1080/14786436908217784}%
  \BibitemOpen
  \bibfield  {author} {\bibinfo {author} {\bibfnamefont {V.}~\bibnamefont
  {Vitek}}\ and\ \bibinfo {author} {\bibfnamefont {F.}~\bibnamefont {Kroupa}},\
  }\href {\doibase 10.1080/14786436908217784} {\bibfield  {journal} {\bibinfo
  {journal} {Philos. Mag.}\ }\textbf {\bibinfo {volume} {19}},\ \bibinfo
  {pages} {265} (\bibinfo {year} {1969})}\BibitemShut {NoStop}%
\bibitem [{\citenamefont {Vitek}\ \emph {et~al.}(1970)\citenamefont {Vitek},
  \citenamefont {Perrin},\ and\ \citenamefont
  {Bowen}}]{doi:10.1080/14786437008238490}%
  \BibitemOpen
  \bibfield  {author} {\bibinfo {author} {\bibfnamefont {V.}~\bibnamefont
  {Vitek}}, \bibinfo {author} {\bibfnamefont {R.~C.}\ \bibnamefont {Perrin}}, \
  and\ \bibinfo {author} {\bibfnamefont {D.~K.}\ \bibnamefont {Bowen}},\ }\href
  {\doibase 10.1080/14786437008238490} {\bibfield  {journal} {\bibinfo
  {journal} {Philos. Mag.}\ }\textbf {\bibinfo {volume} {21}},\ \bibinfo
  {pages} {1049} (\bibinfo {year} {1970})}\BibitemShut {NoStop}%
\bibitem [{\citenamefont {Vitek}(2004)}]{doi:10.1080/14786430310001611644}%
  \BibitemOpen
  \bibfield  {author} {\bibinfo {author} {\bibfnamefont {V.}~\bibnamefont
  {Vitek}},\ }\href {\doibase 10.1080/14786430310001611644} {\bibfield
  {journal} {\bibinfo  {journal} {Philos. Mag.}\ }\textbf {\bibinfo {volume}
  {84}},\ \bibinfo {pages} {415} (\bibinfo {year} {2004})}\BibitemShut
  {NoStop}%
\bibitem [{\citenamefont {E}\ \emph {et~al.}(2002)\citenamefont {E},
  \citenamefont {Ren},\ and\ \citenamefont
  {Vanden-Eijnden}}]{PhysRevB.66.052301}%
  \BibitemOpen
  \bibfield  {author} {\bibinfo {author} {\bibfnamefont {W.}~\bibnamefont {E}},
  \bibinfo {author} {\bibfnamefont {W.}~\bibnamefont {Ren}}, \ and\ \bibinfo
  {author} {\bibfnamefont {E.}~\bibnamefont {Vanden-Eijnden}},\ }\href
  {\doibase 10.1103/PhysRevB.66.052301} {\bibfield  {journal} {\bibinfo
  {journal} {Phys. Rev. B}\ }\textbf {\bibinfo {volume} {66}},\ \bibinfo
  {pages} {052301} (\bibinfo {year} {2002})}\BibitemShut {NoStop}%
\bibitem [{\citenamefont {E}\ \emph {et~al.}(2007)\citenamefont {E},
  \citenamefont {Ren},\ and\ \citenamefont
  {Vanden-Eijnden}}]{:/content/aip/journal/jcp/126/16/10.1063/1.2720838}%
  \BibitemOpen
  \bibfield  {author} {\bibinfo {author} {\bibfnamefont {W.}~\bibnamefont {E}},
  \bibinfo {author} {\bibfnamefont {W.}~\bibnamefont {Ren}}, \ and\ \bibinfo
  {author} {\bibfnamefont {E.}~\bibnamefont {Vanden-Eijnden}},\ }\href
  {\doibase 10.1063/1.2720838} {\bibfield  {journal} {\bibinfo  {journal} {J.
  Chem. Phys.}\ }\textbf {\bibinfo {volume} {126}},\ \bibinfo {eid} {164103}
  (\bibinfo {year} {2007})}\BibitemShut {NoStop}%
\bibitem [{\citenamefont {Ismail-Beigi}\ and\ \citenamefont
  {Arias}(2000)}]{PhysRevLett.84.1499}%
  \BibitemOpen
  \bibfield  {author} {\bibinfo {author} {\bibfnamefont {S.}~\bibnamefont
  {Ismail-Beigi}}\ and\ \bibinfo {author} {\bibfnamefont {T.~A.}\ \bibnamefont
  {Arias}},\ }\href {\doibase 10.1103/PhysRevLett.84.1499} {\bibfield
  {journal} {\bibinfo  {journal} {Phys. Rev. Lett.}\ }\textbf {\bibinfo
  {volume} {84}},\ \bibinfo {pages} {1499} (\bibinfo {year}
  {2000})}\BibitemShut {NoStop}%
\bibitem [{\citenamefont {Segall}\ \emph {et~al.}(2003)\citenamefont {Segall},
  \citenamefont {Strachan}, \citenamefont {Goddard}, \citenamefont
  {Ismail-Beigi},\ and\ \citenamefont {Arias}}]{PhysRevB.68.014104}%
  \BibitemOpen
  \bibfield  {author} {\bibinfo {author} {\bibfnamefont {D.~E.}\ \bibnamefont
  {Segall}}, \bibinfo {author} {\bibfnamefont {A.}~\bibnamefont {Strachan}},
  \bibinfo {author} {\bibfnamefont {W.~A.}\ \bibnamefont {Goddard}}, \bibinfo
  {author} {\bibfnamefont {S.}~\bibnamefont {Ismail-Beigi}}, \ and\ \bibinfo
  {author} {\bibfnamefont {T.~A.}\ \bibnamefont {Arias}},\ }\href {\doibase
  10.1103/PhysRevB.68.014104} {\bibfield  {journal} {\bibinfo  {journal} {Phys.
  Rev. B}\ }\textbf {\bibinfo {volume} {68}},\ \bibinfo {pages} {014104}
  (\bibinfo {year} {2003})}\BibitemShut {NoStop}%
\bibitem [{\citenamefont {Cereceda}\ \emph {et~al.}(2013)\citenamefont
  {Cereceda}, \citenamefont {Stukowski}, \citenamefont {Gilbert}, \citenamefont
  {Queyreau}, \citenamefont {Ventelon}, \citenamefont {Marinica}, \citenamefont
  {Perlado},\ and\ \citenamefont {Marian}}]{0953-8984-25-8-085702}%
  \BibitemOpen
  \bibfield  {author} {\bibinfo {author} {\bibfnamefont {D.}~\bibnamefont
  {Cereceda}}, \bibinfo {author} {\bibfnamefont {A.}~\bibnamefont {Stukowski}},
  \bibinfo {author} {\bibfnamefont {M.~R.}\ \bibnamefont {Gilbert}}, \bibinfo
  {author} {\bibfnamefont {S.}~\bibnamefont {Queyreau}}, \bibinfo {author}
  {\bibfnamefont {L.}~\bibnamefont {Ventelon}}, \bibinfo {author}
  {\bibfnamefont {M.-C.}\ \bibnamefont {Marinica}}, \bibinfo {author}
  {\bibfnamefont {J.~M.}\ \bibnamefont {Perlado}}, \ and\ \bibinfo {author}
  {\bibfnamefont {J.}~\bibnamefont {Marian}},\ }\href {\doibase
  10.1088/0953-8984/25/8/085702} {\bibfield  {journal} {\bibinfo  {journal} {J.
  Phys.: Condens. Matter}\ }\textbf {\bibinfo {volume} {25}},\ \bibinfo {pages}
  {085702} (\bibinfo {year} {2013})}\BibitemShut {NoStop}%
\bibitem [{\citenamefont {Hartley}\ and\ \citenamefont
  {Mishin}(2005)}]{Hartley20051313}%
  \BibitemOpen
  \bibfield  {author} {\bibinfo {author} {\bibfnamefont {C.}~\bibnamefont
  {Hartley}}\ and\ \bibinfo {author} {\bibfnamefont {Y.}~\bibnamefont
  {Mishin}},\ }\href {\doibase 10.1016/j.actamat.2004.11.027} {\bibfield
  {journal} {\bibinfo  {journal} {Acta Mater.}\ }\textbf {\bibinfo {volume}
  {53}},\ \bibinfo {pages} {1313 } (\bibinfo {year} {2005})}\BibitemShut
  {NoStop}%
\bibitem [{\citenamefont {Mendis}\ \emph {et~al.}(2006)\citenamefont {Mendis},
  \citenamefont {Mishin}, \citenamefont {Hartley},\ and\ \citenamefont
  {Hemker}}]{doi:10.1080/14786430600660849}%
  \BibitemOpen
  \bibfield  {author} {\bibinfo {author} {\bibfnamefont {B.~G.}\ \bibnamefont
  {Mendis}}, \bibinfo {author} {\bibfnamefont {Y.}~\bibnamefont {Mishin}},
  \bibinfo {author} {\bibfnamefont {C.~S.}\ \bibnamefont {Hartley}}, \ and\
  \bibinfo {author} {\bibfnamefont {K.~J.}\ \bibnamefont {Hemker}},\ }\href
  {\doibase 10.1080/14786430600660849} {\bibfield  {journal} {\bibinfo
  {journal} {Philos. Mag.}\ }\textbf {\bibinfo {volume} {86}},\ \bibinfo
  {pages} {4607} (\bibinfo {year} {2006})}\BibitemShut {NoStop}%
\end{thebibliography}%

\end{document}